\documentclass[11pt,a4paper]{article}
\usepackage{jheppub}
\usepackage[utf8]{inputenc}
\usepackage[T1]{fontenc}
\usepackage{amsmath}
\usepackage{xcolor}
\usepackage{slashed}
\usepackage{simpler-wick}
\usepackage{soul}
\usepackage{cancel}
\usepackage{subcaption}
\usepackage{graphicx}
\usepackage[normalem]{ulem}
\counterwithin*{equation}{section}

\def\beq{\begin{equation}}
\def\eeq{\end{equation}}
\def\bea{\begin{eqnarray}}
\def\eea{\end{eqnarray}}

\def\D0bar{\overline{D^0}}

\def\ddbar{D^0\overline{D^0}}

\definecolor{darkgreen}{RGB}{0,175,10}
\definecolor{brown}{RGB}{150,50,0}

\title{\textbf{Nonperturbative Dynamics in $D$-meson Mixing}}
\author[a]{Lovro Dulibi\'c,}
\author[a]{Bla\v zenka Meli\'c,}
\author[b]{Alexey A. Petrov}

\affiliation[a]{Ru\dj er Bo\v skovi\'c Institute,
Bĳeni\v cka cesta 32, 10000 Zagreb, Croatia}
\affiliation[b]{Department of Physics and Astronomy, University of South Carolina, Columbia, South Carolina 29208, USA}

\emailAdd{ldulibic@irb.hr}
\emailAdd{melic@irb.hr}
\emailAdd{apetrov@sc.edu}

\date{\today}

\abstract{Theoretical predictions for the $\ddbar$ mixing parameters fall significantly short of experimental measurements, with discrepancies spanning several orders of magnitude. This gap is mainly due to the Glashow–Iliopoulos–Maiani (GIM) mechanism, which suppresses leading-order contributions by high powers of $m_s/m_c$. However, higher-order corrections and nonperturbative effects could reduce this suppression, especially through flavor $SU(3)_F$ symmetry breaking. In this work, we investigate the long-distance contributions from QCD condensates, including, for the first time, the effects of mixed quark-gluon and four-quark condensates. Our results show an increase in the predicted values of $\ddbar$ mixing parameters by two orders of magnitude compared to perturbative NLO result, providing valuable insights into nonperturbative QCD dynamics. Although the theoretical estimates still fall below experimental values, this study represents an important step toward narrowing the gap between theory and observation, highlighting the significance of higher-order $1/m_c$ QCD effects in understanding $\ddbar$ mixing.
}

\begin{document}
\preprint{RBI-ThPhys-2025-35, USC-TH-2025-01}
\maketitle
\flushbottom

%%%%%%%%%%%%%%%%%%%%%%%%%%%%%%%%%%%
\section{Introduction}

The phenomenon of meson-antimeson mixing offers a crucial insight into the structure of fundamental interactions in particle physics and rightfully warrants careful theoretical investigation. Such studies aim to thoroughly understand and distinguish the Standard Model (SM) contributions from potential effects of New Physics (NP) interactions. Despite extensive theoretical and experimental efforts, $D$-meson mixing remains an intriguing puzzle. Unlike well-studied mixing in the $K$, $B$, and $B_s$ meson systems, the Standard Model describes the $D^0$ meson mixing through the interplay of down-type quark dynamics \cite{Petrov:2024ujw,Friday:2025gpj,Lenz:2020awd}. It relies on intricate cancellations between the contributions of the $d$, $s$, and $b$ quarks, making it an excellent area for testing our understanding of the interplay of weak and strong interactions. Experimental measurements of the $\ddbar$ mixing parameters place strong constraints on theoretical models and potential new physics effects. While recent progress at LHCb, Belle II, and other experiments has led to precise measurements of the mixing parameters, the theoretical understanding of the SM contribution to $\ddbar$ mixing remains a challenge. 

Many aspects of this challenge can be understood by analyzing the scales involved in the problem. The effective Hamiltonian formalism requires the separation of these scales. Yet, physically, 
\begin{equation}\label{Intro:scales}
m_q \ll \Lambda \sim m_c \ll m_b,
\end{equation} 
where $m_q$ represents light quark masses, and $\Lambda \sim 1$ GeV parametrizes the typical scale of soft hadronic interactions, which is the scale at which all relevant matrix elements $\langle D^0 | ... |\overline D^0 \rangle$ are calculated. The Glashow-Iliopoulos-Maiani (GIM) mechanism ensures that the effective Hamiltonian describing $\ddbar$ mixing receives contributions proportional to powers of the $d$, $s$, and $b$ quark masses. Among these, the $b$ quark contribution is local at the scale $\Lambda$, but it is always multiplied by the tiny combination of CKM matrix elements, $V_{ub} V_{cb}^*$, which makes this contribution very small (see \cite{Bobrowski:2010xg} for discussion). This contrasts with the situation in studies of mixing of down-type mesons, where the GIM mechanism ensures that the Standard Model contribution is dominated by local effects associated with the top quark.
The remaining contributions from strange and down quarks are nonlocal at the scale $\Lambda$ and mostly cancel each other out because of the GIM mechanism. In fact, their combined contribution is exactly zero in the flavor $SU(3)_F$ limit, where $m_s=m_d$, so the entire effect is an $SU(3)_F$ breaking correction, which makes it important to study.

The light-quark contribution to $\ddbar$ mixing has been studied using a variety of methods, which can roughly be classified as exclusive and inclusive. In exclusive methods, one uses the fact that the mass of the $D$ meson is not very large, so the correlation functions can be saturated by a set of exclusive intermediate states \cite{Golowich:1998pz,Falk:2001hx,Cheng:2024hdo} to obtain the lifetime difference. The mass difference is then obtained using a dispersion relation under various assumptions \cite{Donoghue:1985hh,Falk:2004wg,Li:2022jxc,Li:2020xrz}. The exclusive methods correctly predict the results for the width and mass difference parameters in the $\ddbar$ system within an order magnitude and requires observations of decay rates with per-mil precision, which is currently not available \cite{HeavyFlavorAveragingGroupHFLAV:2024ctg}. The inclusive methods attempt to compute the mixing parameters in QCD using quark degrees of freedom \cite{Datta:1984jx,Golowich:2005pt,Georgi:1992as,Bigi:2000wn,Bobrowski:2010xg,Bobrowski:2012jf}, computing the nonlocal contribution due to the light quarks using the methods of Operator Product Expansion (OPE). The problem can also be approached via model-dependent methods \cite{Xie:2025hzx}. Recently, lattice QCD approaches are being developed to address the problem \cite{DiCarlo:2025mvt} as well. It was demonstrated in \cite{Lenz:2020efu} that $\ddbar$ mixing could be resolved perturbatively where $SU(3)_F$ breaking is introduced at the leading order (LO) in $\alpha_s$ by selecting the renormalization scale for each internal quark contribution separately.

The nonlocality of the light-quark contribution can be studied in the heavy-quark limit, where one {\it assumes} that $\Lambda \ll m_c$. Although it may seem contrary to Eq.~(\ref{Intro:scales}), it enables us to use the powerful method of effective Hamiltonians for this problem and expand the nonlocal contributions from light quarks in powers of $m_q/m_c$ and $\Lambda/m_c$, resulting in a series of matrix elements of local operators. The hope is that computing enough terms in this expansion will help us understand what happens in the realistic case $\Lambda \simeq m_c$.

This paper offers a comprehensive analysis of higher-order $1/m_c$ contributions to $\ddbar$ mixing by analyzing the contributions stemming from the nonlocal QCD condensates. It is organized as follows. We set up the calculation in Section \ref{GC} and define and review the leading-order results for the mixing parameter $x_D$ in Section \ref{sec:LO}. We explicitly compute the higher-order $1/m_c$ effects of QCD condensates in Section \ref{sec:HigherDimOp}, focusing on terms that contribute with fewer powers of $m_s$. The numerical results are presented in Section \ref{sec:Summary}, and we conclude in Section \ref{sec:Conclusions}.

%%%%%%%%%%%%%%%%%%%%%%%%%%%%%%%%%%%%%%%%%%%%%%
\section{General comments}\label{GC}

The complexity of theoretical studies of the mixing phenomenon in charmed mesons arises from the fact that the mass of the charmed quark—and consequently the charmed meson states—while clearly larger than $\Lambda_{\rm QCD} \sim 400$ MeV, falls within the region populated by light-meson resonances. This complicates the analysis of QCD dynamics and provides a theoretical laboratory for evaluating the applicability of heavy-quark methods. 

The mixing of $D^0$ and $\overline D^0$ mesons occurs due to the presence of $\Delta C=2$ interactions, which introduce off-diagonal pieces in the $2\times 2$ effective Hamiltonian $\left (M - i \Gamma/2\right)$ that describes the combined propagation of those states,
\beq
i \frac{\partial}{\partial t} | D\rangle = \left (M - \frac{i}{2}\, \Gamma\right) | D\rangle \,,
\eeq
where $| D\rangle$ is the vector of the $D^0$ and $\overline D^0$ states. The off-diagonal terms can be symbolically written as an expansion, 
\beq\label{M12}
\left (M - \frac{i}{2}\, \Gamma\right)_{12} =
  \frac{1}{2M_D}\, \langle \overline D^0 | {\cal H}_w^{\Delta C=2} | D^0 \rangle +
  \frac{1}{2M_D}\, \sum_n \frac{\langle \overline D^0 | {\cal H}_w^{\Delta C=1} | n 
  \rangle\, \langle n | {\cal H}_w^{\Delta C=1} | D^0 \rangle}{M_D-E_n+i\epsilon} \ \ ,
\eeq
where ${\cal H}_w^{\Delta C=2}$ and ${\cal H}_w^{\Delta C=1}$ are the effective $\Delta C=2$ and $\Delta C=1$ Hamiltonians, respectively. Both effective Hamiltonians can be generated in the SM from 
\bea\label{Heff}
{\cal H}_w^{\Delta C=1} &=& \frac{G_F}{\sqrt{2}} 
\sum_{q,q'}\ V_{cq}^*V_{uq'} \left[ C_1(\mu) Q_1 + C_2(\mu) Q_2 \right]\ \,,
\eea 
with 
\bea
Q_1 \ &=& \ \left(\overline{q}^i \gamma^\mu(1-\gamma^5)c^i\right)\,
\left(\overline{u}^j \gamma^\mu(1-\gamma^5)q'^j\right) \,,
\nonumber \\
Q_2 \ &=& \ \left(\overline{q}^i\gamma^\mu(1-\gamma^5) c^j\right)\,
\left(\overline{u}^j \gamma^\mu(1-\gamma^5)q'^i\right)\,, \qquad { q,q' = d,s,b} \,.
\nonumber
\eea 
The $C_{1,2}(\mu)$ in Eq.~(\ref{Heff}) are the perturbative Wilson coefficients evaluated at the scale $\mu$ \cite{Buchalla:1995vs}.
The ${\cal H}_w^{\Delta C=2}$ is obtained by integrating out modes with momenta $p\gg m_c$ from the $T$-product of two ${\cal H}_w^{\Delta C=1}$  effective Hamiltonians. The $b$-quark contribution is always multiplied by a small CKM factor $|V_{ub} V_{cb}^*|^2/|V_{us} V_{cs}^*|^2 = {\cal O}(10^{-6})$, so we will neglect it from now on. However, the effective Hamiltonian ${\cal H}_w^{\Delta C=2}$ could receive potentially large contributions from some heavy NP states. Since it is generated by the heavy states, being SM or not, it cannot significantly contribute to the $\Gamma_{12}$ part of the mixing matrix and thus, to the lifetime difference (see, however, \cite{Golowich:2006gq}) and therefore can be completely neglected. 

It is convenient to diagonalize the mass matrix $\left (M - i  \Gamma/2\right)$ to determine the propagating mass eigenstates $D_L$ and $D_S$,  which are superpositions of the flavor eigenstates,
\begin{equation} \label{definition1}
   | D_{L,S} \rangle = p\, | D^0 \rangle \pm q\, | \overline D^0 \rangle\ \ ,
\end{equation}
where $|p|^2 + |q|^2=1$. Neglecting CP-violation In the SM, known to be small, we have $p=q$, so $| D_{L,S} \rangle$ become the CP eigenstates $| D_{\pm} \rangle$ with  ${\cal C}{\cal P} | D_{\pm} \rangle = \pm | D_{\pm} \rangle$. We can then define the mass and the width differences, 
\begin{equation} 
\Delta M_{\rm D} \equiv M_{D_+} - M_{D_-} 
\qquad  {\rm and} \qquad \Delta \Gamma_{\rm D} \equiv \Gamma_{D_+} - 
\Gamma_{D_-} \ \ .
\label{diffs}
\end{equation}
Normalizing them to the average width of the two neutral $D$ meson mass eigenstates $\Gamma_{\rm D}$, the two relevant $\ddbar$ mixing parameters,  $x_D$ and $y_D$, are defined as:
\beq
x_D \equiv \frac{\Delta M_{\rm D}}{\Gamma_{\rm D}}, \qquad 
y_D \equiv \frac{\Delta \Gamma_{\rm D}}{2\Gamma_{\rm D}}.
\eeq
In this paper, we will calculate the mass difference $x_D$ using heavy quark expansion methods \cite{Petrov:2016azi}, with the introduction of the nonperturbative QCD effects from the QCD condensates. We will treat the charm quark as heavy, assuming that $m_c > \Lambda$, but {\it not} $m_c \gg \Lambda$, where $\Lambda \sim 1$ GeV is the hadronic scale. This is a standard assumption in quark-level calculations of mass and lifetime differences of $D$-mesons. 

Our calculation employs an operator product expansion (OPE) \cite{Georgi:1992as,Bigi:2000wn,Bobrowski:2010xg}. In the limit where $m_c > \Lambda$, the momentum flowing through the light degrees of freedom in the intermediate state is large. Therefore, an OPE is performed by expanding the second term in Eq.~(\ref{M12}) into a series of matrix elements of local operators. 

%%%%%%%%%%%%%%%%%%%%%%%%%%%%%%%%%%%
\section{The leading order computation}
\label{sec:LO}

The leading-order result for the mass and lifetime differences of $D$-mesons in $1/m_c$ expansion is well known \cite{Datta:1984jx,Golowich:2005pt,Georgi:1992as}. We will reproduce it here to demonstrate the heavy-mass expansion technique and set up further computation. 

\begin{figure}
    \centering
    \includegraphics[width=0.5\linewidth]{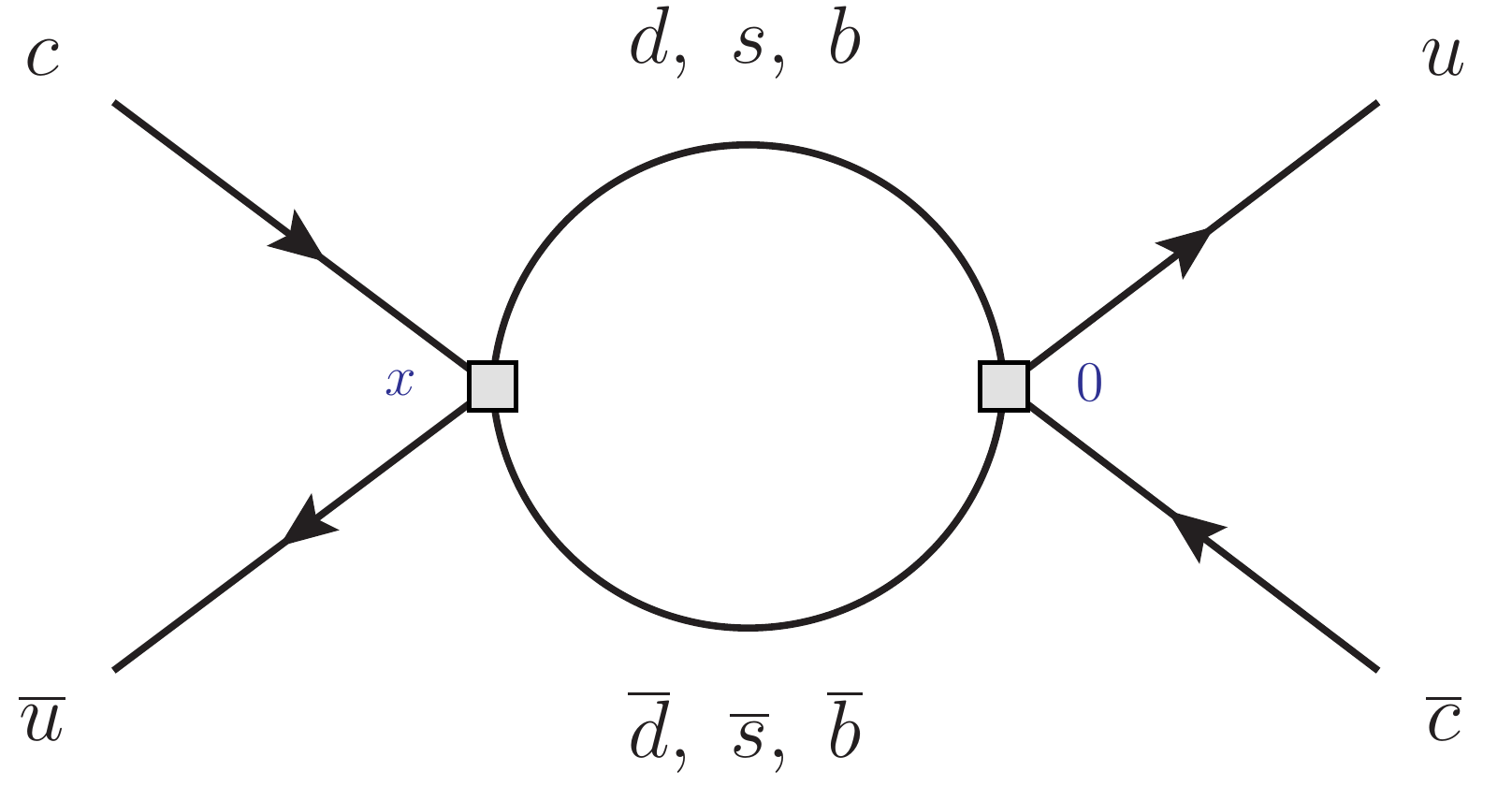}
    \caption{Diagram of the dimension-6, leading order perturbative contribution. The gray squares denote weak effective vertices in which local $Q_1$ and $Q_2$ operators are inserted.}
    \label{fig:boxdiag}
\end{figure}

The possibility of using the OPE in this calculation stems from the fact that the charm quark injects large, ${\cal O}(m_c)$, momentum into the intermediate quark system \cite{Georgi:1992as}. Expanding the correlation function representing a time-ordered product of effective operators $O_{1,2}$, we can contract the light-quark fields as\footnote{In intermediate steps of the calculation color indices of quark fields are left general and contractions of color singlet and non-singlet operators $Q_1,\, Q_2$ are introduced later.} (see Fig. \ref{fig:boxdiag})
\begin{equation}\label{eq:MIX}
\wick{ T_{q q'}^{(6)} = \frac{i G_F^2}{2}
\int d^4 x \ \big(\overline{c}^i(x) \Gamma^\mu \c1{q}^j(x) \big)\big(\c2{\overline{q'}}^k(x) \Gamma_\mu u^l (x)\big)
\big(\overline{c}^m(0) \Gamma^\nu  \c2 q'^{\,n} (0) \big)\big(\c1{\overline{q}}^p(0) \Gamma_\nu u^q(0)}\big),
\end{equation}
where the quark field contractions are given by the propagators, 
\begin{equation}
    \wick{
    \c \psi(x)_\alpha^a \c{\overline{\psi}}(y)_\beta^b = \delta^{ab}\int \frac{d^4k}{(2\pi)^4}i\frac{(\slashed{k}+m)_{\alpha\beta}}{k^2-m^2}e^{-ik(x-y)}
    },
\end{equation}
with the Greek indices $\alpha$ and $\beta$ representing the spinor indices, and the Latin indices $a$ and $b$ representing the color indices. Changing the order of integration,  
\begin{eqnarray}
T_{q q'}^{(6)} = \frac{i G_F^2}{2} \delta^{jp}\delta^{nk}
\int\frac{d^4l}{(2\pi)^4}\frac{i}{k^2-m_q^2}\frac{i}{l^2-m_{q'}^2} \int d^4x\, e^{i(p_c -k+l+p_u)}\int \frac{d^4k}{(2\pi)^4}
\nonumber \\
 \times \Big( \overline{c}^i\Gamma^\mu(\slashed{k}+m_q)\Gamma^\nu u^q\Big)
 \left( \overline{c}^m\Gamma_\nu (\slashed{l}+m_{q'})\Gamma_\mu u^l \right), 
\end{eqnarray}
where the momenta of the internal quarks are $k$ and $l$, and we rephased the $c$ and $u$ quark fields, e.g., $c(x)=e^{ip_c x}c(0)$. We also simplify notation and write $q$ instead of $q(0)$ since all fields are now evaluated at the same point. Integrating over $x$ and $l$ we get
\begin{equation}
    T_{q q'}^{(6)}=-\frac{i G_F^2}{2} \delta^{jp}\delta^{nk}\int\frac{d^4 k}{(2\pi)^4}\frac{k^\alpha(k^\beta-p^\beta)}{(k^2-m_{q}^2)((k-p)^2-m_{q'}^2)} \Big(\overline{c}^i\Gamma^\mu\gamma_\alpha\Gamma^\nu u^q\Big)\Big(\overline{c}^m\Gamma_\nu\gamma_\beta\Gamma_\mu u^l\Big).
\end{equation}
Plugging in the color structure of effective operators $Q_1$ and $Q_2$, and decomposing the tensor integral we get
\begin{eqnarray}\label{eq:LO_intermediate-stepXY}
    T_{q q'}^{(6)}  =\frac{i 8 G_F^2}{2}&\bigg[&\left( ~2X(m_{q},m_{q'},p^2) C_1^2+(2X(m_{q},m_{q'},p^2) \right.
    \\
 &+&     \left. Y(m_{q},m_{q'},p^2) m_c^2)(2C_1C_2+N_cC_2^2)\right)O_{V-A}
    \nonumber \\
    &-&2Y(m_{q},m_{q'},p^2) m_c^2(C_1^2-2C_1C_2-N_cC_2^2)O_{S-P} \bigg],
    \nonumber
\end{eqnarray}
where the functions $X(m_{q},m_{q'},p^2)$ and $Y(m_{q},m_{q'},p^2)$ are listed in appendix \ref{app:integral}, and the operators are defined as 
\begin{eqnarray}
O_{V-A} &=& \left(\overline{c} \gamma^\mu (1-\gamma^5)u\right)\left(\overline{c}\gamma_\mu(1-\gamma^5)u\right),
\nonumber \\
O_{S-P} &=& \left(\overline{c} \left(1-\gamma^5\right) u\right)\left(\overline{c} \left(1-\gamma^5\right) u\right).
\end{eqnarray}
The matrix elements of those four-quark operators can be computed on the lattice or with other nonperturbative QCD methods. The decay width and mass difference can be obtained as absorptive (imaginary) and dispersive (real) parts of the sum of diagrams with the relevant internal quarks. Taking the appropriate matrix elements, we obtain
\begin{eqnarray}\label{eq:xandywrittenout}
        x_D^{(6)} = \frac{G_F^2}{2}\frac{1}{2M_D\Gamma_D}\mathrm{Re}\Big[&&\xi_s^2
        \langle D^0 | (T_{ss}^{(6)}-2T_{sd}^{(6)}+T_{dd}^{(6)}) |\overline D^0\rangle 
         \\
        &+& 2\xi_s\xi_b \langle D^0 |(T_{bs}^{(6)}-T_{bd}^{(6)}-T_{sd}^{(6)}+T_{dd}^{(6)})|\overline D^0\rangle 
        \nonumber \\
        &+&\xi_b^2 \langle D^0 |(T_{bb}^{(6)}-2T_{bd}^{(6)}+T_{dd}^{(6)})|\overline D^0\rangle\Big],
        \nonumber\\
        y_D^{(6)} =\frac{G_F^2}{2}\frac{1}{2M_D\Gamma_D} \mathrm{Im}\Big[ &-&\xi_s^2 \langle D^0 |(T_{ss}^{(6)}-2T_{sd}^{(6)}+T_{dd}^{(6)})|\overline D^0\rangle
       \\
        &+& 2\xi_s\xi_b\langle D^0 | (T_{sd}^{(6)}-T_{dd}^{(6)})|\overline D^0\rangle
        -\xi_b^2 \langle D^0 |T_{dd}^{(6)}|\overline D^0\rangle \Big],
        \nonumber
\end{eqnarray}
where the CKM factors are $\xi_q=V_{cq}V^*_{uq}$. Neglecting small terms proportional to $\xi_b$ \cite{Bobrowski:2010xg,Golowich:2005pt}, and defining $x_s = m_s/m_c$, we find that the leading order result for $x_D^{(6)}$ scales as ${\cal O}(x_s^4)$, 
\beq
x_D^{(6)}=-\frac{G_F^2}{2}\frac{ m_c^2}{2M_D\Gamma_D} \xi_s^2 x_s^4 \frac{1}{2\pi^2}\left[C_1^2 \langle O_{V-A} \rangle + 2 \left(C_1^2 -2 C_1C_2 - 3 C_2^2\right) \langle O_{S-P} \rangle \right].
\eeq
Factorizing the product of currents in the operators and introducing the so-called $B$-bag model parameters, to parameterize the deviation from the factorization via the bag model, the matrix elements of the operators can be written as
\beq
\langle \D0bar | O_{V-A} | D^0 \rangle \, = \frac{8}{3}f_D^2 M_D^2 B_{\rm D}
\qquad {\rm and} \qquad 
\langle \D0bar | O_{S-P} | D^0 \rangle \, = -\frac{5}{3}f_D^2 M_D^2
{\overline B}_{\rm D}^{(S)} \ \ ,
\label{b-fctr}
\eeq
where ${\overline B}_{\rm D}^{(S)} \equiv B_{\rm D}^{(S)}M_D^2/m_c^2$. 
Writing out the matrix elements using Eq.~(\ref{b-fctr}), we finally get
\begin{eqnarray}\label{xLO}
x_D^{(6)} &=& 
-\frac{G_F^2 m_c^2 f_D^2 M_D}{3 \pi^2 \Gamma_D} ~\xi_s^2~
x_s^4~ \bigg[  C_1^2 B_{\rm D} - \frac{5}{4} (C_2^2 - 
2 C_1 C_2 - 3 C_2^2) {\overline B}_{\rm D}^{(S)}\bigg].
\end{eqnarray}
Similarly, for the lifetime difference, the leading order result for the $y_D^{(6)}$ is proportional to ${\cal O}(x_s^6)$,
\begin{eqnarray}\label{yLO}
y_D^{(6)} = -\frac{G_F^2 m_c^2 f_D^2 M_D}{3 \pi \Gamma_D} ~\xi_s^2~
x_s^6~ \left( C_1^2 - 2 C_1 C_2 - 3 C_2^2 \right) \left[ 
B_{\rm D} - \frac{5}{2} {\overline B}_{\rm D}^{(S)}
\right]. 
\end{eqnarray}
The most important lesson from calculating the leading-order results is the fact that the matrix elements of the dimension-six operators are suppressed by high powers of $m_s$. As it was shown that the leading order corrections to the symmetry limit scale as ${\cal O}(m_s^2)$ \cite{Falk:2001hx}, one might wonder if those corrections could appear among higher-order terms in $1/m_c$ expansion.

Curiously, it was noticed that higher order corrections in $\alpha_s$ can soften the dependence on the light quark mass: the one-loop QCD corrections make the $y_D \sim {\cal O} (x_s^4)$ \cite{Golowich:2005pt}. Similarly, some higher order terms in $1/m_c$ expansion can lead to a parametrically lower $m_s$-dependence of $x_D$ \cite{Georgi:1992as,Ohl:1992sr}. In our heavy quark limit, with $m_c\ge \Lambda$, but not $m_c\gg \Lambda$, higher order suppression of $\Lambda/m_c$ is numerically not as important as $m_s/m_c$ suppression \cite{Bigi:2000wn}. We will identify and consider those (technically) subleading terms in the next section.

%%%%%%%%%%%%%%%%%%%%%%%%%%%%%%%%%%%
\section{Higher-dimensional operator contributions from QCD condensates}\label{sec:HigherDimOp}

To understand the importance of higher-order $1/m_c$ contributions to $x_D$, it is helpful to clarify the structure of the result obtained earlier. As shown in Section \ref{sec:LO}, the main contribution to the mixing parameter $x_D$ scales with the fourth power of the light quark mass. This can be simply explained: the GIM mechanism ensures the cancellations of the diagrams that do not depend on the internal quark masses. Inserting the quark-mass operator causes a chirality flip on each internal quark line, turning the left-handed strange quarks into right-handed ones. Since $W$-bosons only couple to the left-handed quarks, another mass insertion is required to flip the quark's chirality again, leading to a suppression of $x_D\sim m_s^4$ for the leading-order term in the $1/m_c$ expansion. 

It was pointed out \cite{Georgi:1992as,Ohl:1992sr,Bigi:2000wn} that the higher-order effects caused by local quark condensates can also induce chirality flips without factors associated with the small quark masses. Such effects appear at higher orders in $\Lambda/m_c$, but since $\Lambda \simeq m_c$, those terms might give a dominant contribution to the mixing parameters, leading to a combined expansion in $m_q^n \Lambda^m/m_c^{n+m}$. The previous studies argued that if $m_s \ll \Lambda$, the chirality flip would be associated with the larger scale $\Lambda$, reducing the dependence on $m_s$. As $m_c>\Lambda$, but not $m_c \gg\Lambda$, the suppression by powers of $m_c$ is less effective than the reduction of powers of $m_s$. This artificially enhances the parametrically $1/m_c$-suppressed terms, making them numerically leading.
In the following sections, we will explain how these terms emerge in the $1/m_c$ expansion of the correlation function $T_{q q'}$. Our approach will focus only on the higher-order terms in $1/m_c$ that result in fewer powers of $m_s$.

\begin{figure}
    \centering
    \includegraphics[width=1.\linewidth]{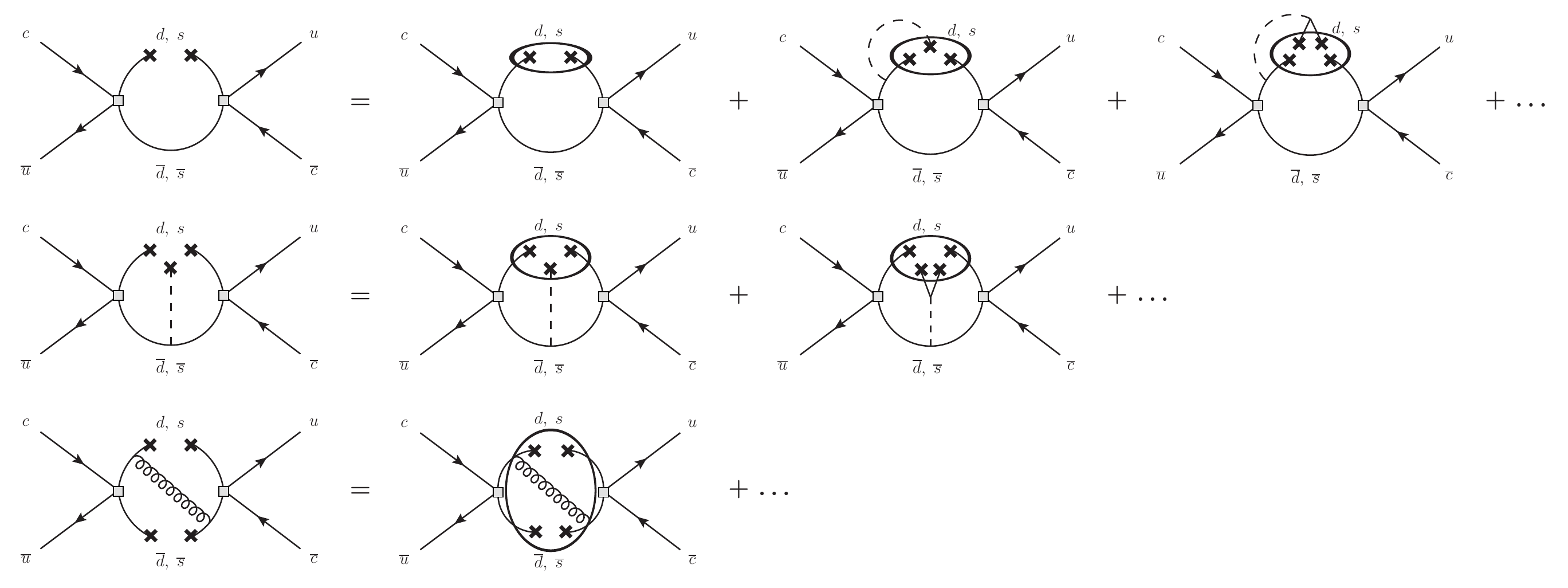}
   \caption{Diagrammatic representation of the local expansions of nonlocal QCD condensates. Ovals denote local condensates. The separate rows correspond to \eqref{eq:qq_cond}, \eqref{eq:GqqExpansion}, and \eqref{eq:4qfactorization}, respectively. Note, for the case of quark-quark and mixed quark-gluon condensates, at both, the upper and the lower line the condensates can be formed.}
    \label{fig:all_condensates}
\end{figure}

We will describe the higher-order effects by computing nonperturbative  contributions from nonlocal QCD condensates, up to dimension-6, i.e. up to dimension-12 operators. By using the background field method, the nonlocal condensates will be expressed in terms of local condensates in coordinate space \cite{pascual2014qcd, Bagan:1985zp,Bagan:1993by,Yndurain:1989hp,Generalis:1983hb,Gromes:1982su,Grozin:1994hd}. Schematically, the expansion is sketched in Fig. \ref{fig:all_condensates}.

%%%%%%%%%%%%%%%%%%%%%%%%%%%%%%%%%%%%%%%%%%
\subsection{Dimension-nine operator contributions}
\label{sec:qq}

\begin{figure}[b]
    \centering
    \includegraphics[width=0.5\linewidth]{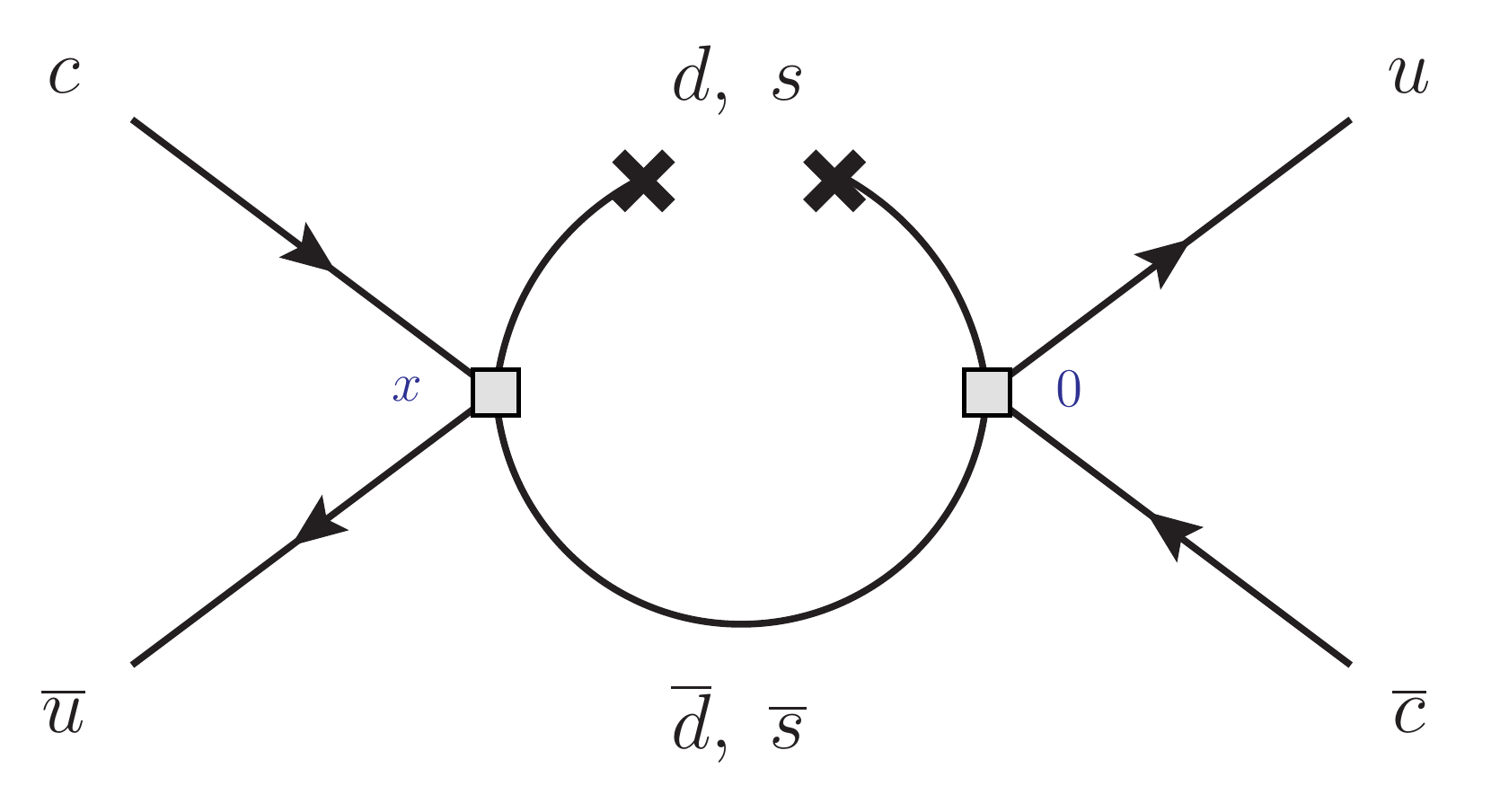}
    \caption{Diagram of the dimension-9 contribution proportional to the quark-quark condensates, $\langle\overline{q}(x) q(0)\rangle$. The same contribution will appear also at the bottom internal line.}
    \label{fig:qq_condensate}
\end{figure}

As we saw in the previous section, the GIM mechanism cancels out the high-momentum parts of the relevant diagrams, enhancing the dependence of the correlation function $T_{q q'}$ on the low-energy QCD dynamics. The $1/m_c$ expansion of $T_{q q'}$ contains a multitude of terms. We will identify those that provide the dominant effect at each order in $1/m_c$, up to dimension-12 operator contributions. Such terms will have the fewest factors of $m_s$.

It is clear that simple derivative insertions do not alter the power of the strange quark mass in the expressions for the calculated contribution. As was pointed out \cite {Georgi:1992as,Bigi:2000wn}, it is the diagrams involving unconnected quark pairs that can potentially lead to the enhanced contributions.

The parametrically dominant non-perturbative contribution arises from a quark pair that we leave uncontracted as depicted in figure \ref{fig:qq_condensate}. This corresponds to the following term
\begin{equation}\label{Dim9Exp}
\wick{T_{q q'}^{(9)} =\frac{iG_F^2}{2} \int d^4 x\big(\overline{c}^i(x) \Gamma_\mu \c{q'}^j(x) \big)\big(\overline{q}^{k}(x) \Gamma^\mu u^l(x)\big)\big(\overline{c}^m(0) \Gamma_\nu q^n(0)\big)\big(\c{\overline{q'}}^p(0) \Gamma^\nu u^q(0)\big)},
\end{equation}
where we denote the uncontracted (which will form the condensate) and propagated quark fields as $q$ and $q'$, respectively. Thus, the mass difference of six-quark, dimension-nine operators will be proportional to the following matrix elements,
\beq
x_D^{(9)} \sim \sum_{q=s,d} C_q \langle D^0 | \left(\overline{c} \Gamma_1 u\right) \left(\overline{q} q\right)\left(\overline{c}\Gamma_2 u\right) |\overline D^0\rangle,
\eeq
where $C_q$ are the coefficients that depend on $G_F$, masses, and CKM factors. Each matrix element of a dimension-nine operator can be calculated using the factorization approximation, which is implemented by inserting vacuum states, so that 
\beq
\langle D^0 | \left(\overline{c} \Gamma_1 u\right) \left(\overline{q} q\right) \left(\overline{c}\Gamma_2 u\right) |\overline D^0\rangle \sim \langle\overline{q} q\rangle  \langle D^0 | \left(\overline{c} \Gamma_1 u\right) \left(\overline{c}\Gamma_2 u\right) |\overline D^0\rangle.
\eeq
This results in a contribution proportional to a nonlocal vacuum quark condensate $\langle\overline{q} q\rangle$. 
%\BM{The previous studies argued that if $m_s \ll \Lambda$, the chirality flip would be associated with the larger scale $\Lambda$, reducing the dependence on $m_s$. As $m_c>\Lambda$, but not $m_c \gg\Lambda$, the suppression by powers of $m_c$ is less effective than the reduction of powers of $m_s$. This artificially enhances the parametrically $1/m_c$-suppressed terms, making them numerically leading.}
%\bmC{this part is somehow not belonging here, we do not discuss any $m_s$ dependence yet...suggestions?}

Since we will be employing the factorization ansatz, we write out the spinor indices ($\alpha$, $\beta$) explicitly to isolate the quark pair which forms the quark condensate,
\begin{equation}\label{Correlator_dim9}
T_{q q'}^{(9)}=\frac{i G_F^2}{2} \int d^4x \big(\overline{c}^i(x) \Gamma_\mu S(x, 0)^{j p} \Gamma^\nu u^q(0)\big)\big(\overline{c}^m(0) \Gamma_\nu\big)_\beta\big(q^n(0)\big)_\beta\big(\overline{q}^k(x)\big)_\alpha\big(\Gamma^\mu u^l(x)\big)_\alpha.
\end{equation}
The remaining propagator can further be expanded in a background gluon field as in \cite{Novikov:1984ecy,Piscopo:2021ogu},
\beq\label{PropExp}
S(x,0)=S_0(x,0)+S_1(x,0)+\ldots\, .
\eeq
The free propagator $S_0(x,0)$ forms the terms with a quark-quark condensate, while the background field correction to the propagator $S_1(x,0)$ forms the quark-gluon-quark (mixed) condensate which we calculate in section \ref{sec:Gqq}.

The calculation can, in fact, be performed directly in coordinate space. The light quark propagator is
\begin{equation}\label{QuarkProp}
S_0(x, 0)^{jp}=\delta^{jp}\left[-\frac{i}{4 \pi^2} \frac{m_{q'}^2}{x^2} K_2\left(m_{q'} \sqrt{-x^2}\right) \slashed{x}+\frac{1}{4 \pi^2} \frac{m_{q'}^2}{\sqrt{-x^2}} K_1\left(m_{q'} \sqrt{-x^2}\right)\right],
\end{equation}
where the $K_{1,2}$ are the modified Bessel functions. Additionally, we can write the nonlocal condensates as an expansion in terms of local condensates (for a comprehensive overview, see \cite{pascual2014qcd,Ioffe:2002ee}). For the nonlocal two-quark condensate, the expansion, up to the dimension-6 contributions, reads \cite{pascual2014qcd} 
\begin{eqnarray}\label{eq:qq_cond}
\langle\overline{q}^i(x)q^j(0)\rangle &=& \delta^{ij}\frac{\langle\overline{q}q\rangle_0}{12}\Biggl[\delta_{\alpha\beta}\left(1+\frac{x^2}{8}\lambda^2_q+\ldots\right) \nonumber \\
&+& i\slashed{x}_{\beta\alpha}\left(\frac{m}{4}+\frac{x^2}{4}\left(\frac{m\lambda_q^2}{12}-\frac{2}{81}\pi\alpha_s^{NP}\frac{\langle\overline{q}q\rangle_0^2}{\langle\overline{q}q\rangle_0}\right)+\ldots\right)\Biggr].
\end{eqnarray}
The expansion introduces higher-dimensional condensates, of dimension-5 - the mixed quark-gluon condensate $\langle \overline{q} \sigma G q \rangle$, and dimension-6 - the four-quark condensate contributions $\langle \overline{q} q \overline{q} q \rangle$, which local contributions are denoted above in the factorization approximation as $\lambda_q^2 \langle\overline{q}q\rangle_0$ and $\langle\overline{q}q\rangle_0^2$, respectively. The parameter $\lambda_q^2$ will be introduced in Eq.~(\ref{MixCondParam}) and we take $\alpha_s^{NP}=1$.
To get the contribution from the dimension-9 operators proportional to the nonlocal quark-quark condensates, we insert Eq.~(\ref{eq:qq_cond}) and Eq.~(\ref{QuarkProp}) into the correlation function of Eq.~(\ref{Correlator_dim9}). Looking at the Dirac structure, we find only the term proportional to $\slashed{x}$ survives. Note that the nonvanishing piece comes exclusively from the nonlocality of $\langle\overline{q}^i(x)q^j(0)\rangle$.

Evaluating the integral, we arrive at the contribution to $x_D$ that is proportional to the quark condensate insertion. Before presenting the answer, it might be useful to discuss the structure of the result. Putting all contributions together leads to
\beq\label{xqq}
x_D^{(9)}\propto \left(x_s^2 -x_d^2\right) \left (x_s \langle\overline{s}s\rangle_0 - x_d \langle\overline{d}d\rangle_0 \right )+\mathcal{O}(x^4_s,x^4_d )
,
\eeq
As $x_d \ll x_s$ it is convenient to set $x_d=0$, and keep the leading order terms in $x_s$, 
\beq\label{xqq_lead}
x_D^{(9)} \, {\propto} \, x_s^3 \langle\overline{s}s\rangle_0   \,.
\eeq
The dimension-nine operator contribution to the mixing parameter $x_D$ is indeed proportional to $x_s^3$, i.e., contains one less power of $m_s$ than the dimension-six contribution in Eq.~(\ref{yLO}). At this order in $x_s$, there are no higher-order condensate contributions, and therefore this leading contribution is proportional to the $\langle\overline{s}s\rangle_0$ only, i.e. the mixed quark-gluon condensate and the four-quark condensate contributions from (\ref{eq:qq_cond}) are not present. 
The full result, up to the leading power on $m_s/m_c$ is then
\begin{equation}\label{AnsDim9}
    x_D^{(9)} =\frac{G_F^2}{3}\frac{\xi_s^2 m_c^2}{2M_D\Gamma_D}\frac{\langle\overline{s}s\rangle_0}{m_c^3} \, x_s^3 \Big[\left(C_1^2-2C_1C_2-3C_2^2 \right)\left(\langle O_{V-A} \rangle+4\langle O_{S-P} \rangle\right) \Big],
\end{equation}
which gives explicitly the leading contribution to $x_D$ from the dimension-nine operators proportional to the quark-quark condensates, obtained in the factorization approximation. We note that while it is historically taken that $\langle \overline q q\rangle_0 \sim {\cal O} (1~\text{GeV}^3)$, the recently extracted value using lattice QCD or QCD sum rules is numerically lower. We will discuss the numerical value of this contribution in Section \ref{sec:Summary}.

%%%%%%%%%%%%%%%%%%%%%%%%%%%%%%%%%%%%%%%%%%
\subsection{Dimension-eleven operator contributions}
\label{sec:Gqq}
\begin{figure}
    \centering
    \includegraphics[width=0.5\linewidth]{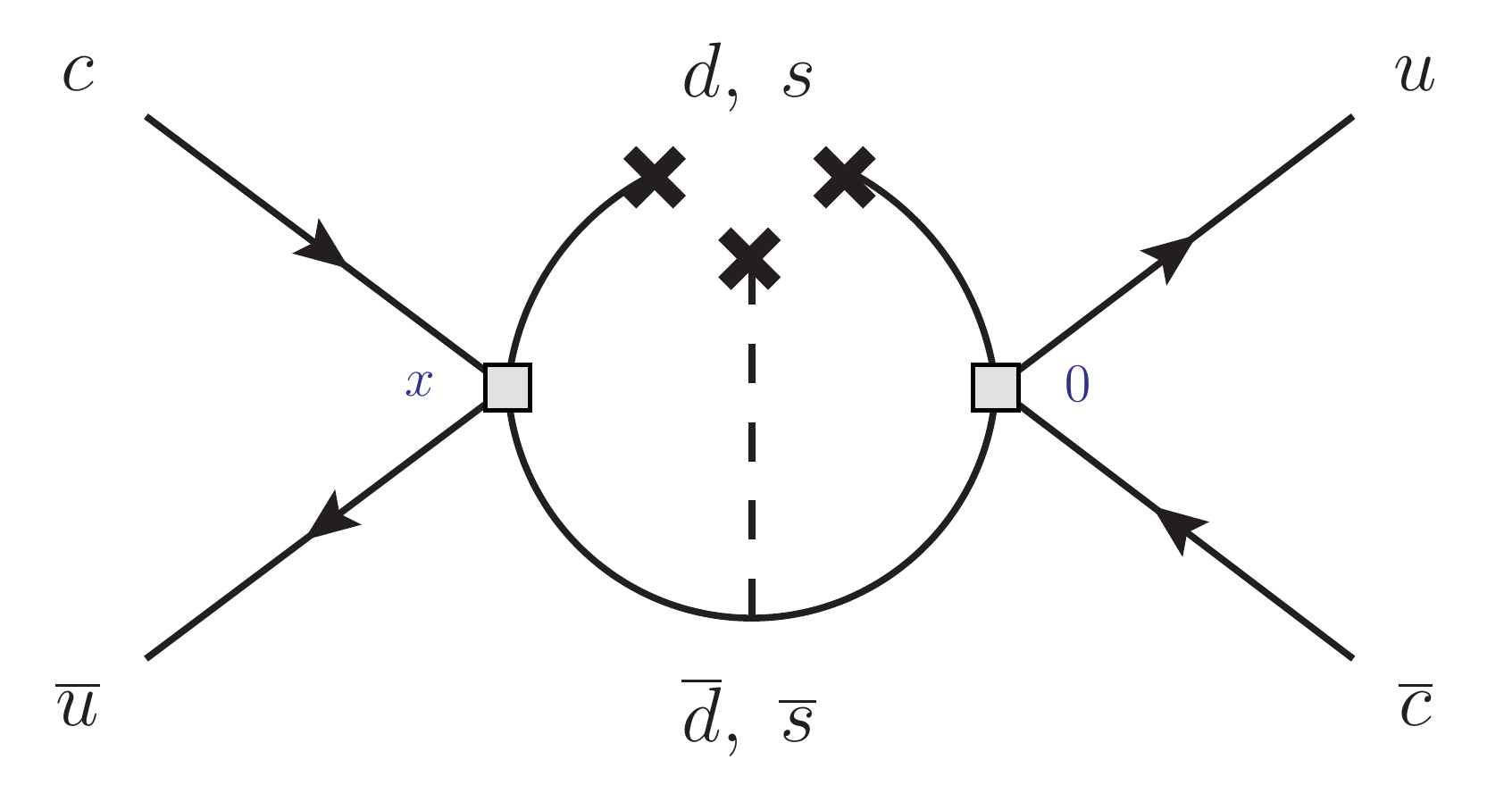}
    \caption{Diagram of the dimension-11 contribution proportional to the mixed quark-gluon condensates, $\langle\overline{q}(x) \sigma G q(0)\rangle$. The same contribution will appear also at the bottom internal line.}
    \label{fig:mixed_condensate}
\end{figure}
It is interesting to extend the computation to even higher orders in the operator product expansion. At the order of dimension-eleven operators one encounters a contribution from the nonlocal mixed quark-gluon condensate, 
\beq
\langle\overline{q} i g \sigma  G q \rangle \equiv
\langle\overline{q} i g \sigma^{\mu\nu} G_{\mu\nu}^a \frac{\lambda^a}{2} q \rangle,
\eeq
where $\lambda^a$ are the Gell-Mann matrices. For the local mixed quark-gluon condensate it is often convenient to define 
\beq\label{MixCondParam}
\frac{\langle\overline{q} ig\sigma^{\mu\nu}G_{\mu\nu} q \rangle_0}{\langle\overline{q} q \rangle_0} =
m_0^2 \equiv 2\lambda_q^2+2m_q^2, 
\eeq
where the parameter $m_0^2$ is computed using non-perturbative methods, such as QCD sum rules or lattice QCD. The Eq.~(\ref{MixCondParam}) can be used as a definition for the parameter $\lambda_q^2$, for which we use the often quoted value $\lambda_q^2=(0.4\pm0.1)$ GeV$^2$ \cite{Belyaev:1982sa}.\footnote{Note that in principle this value could differ for light quarks and the strange quark. However, due to lack of precise analyses we assume $\lambda_s^2=\lambda_d^2$. For more details see discussion in \cite{Gubler:2018ctz}.} The mixed quark-gluon condensate is also a chiral order parameter of QCD, as it flips the chirality of the light quark,
\beq
\langle\overline{q} \sigma G q \rangle = \langle\overline{q_R} \sigma G  q_L \rangle + \langle\overline{q_L} \sigma G  q_R \rangle,
\eeq
which produces the same effect as the quark condensate, removing a power of the light-quark mass in $x_D$,  discussed in the previous section. 

The mass difference will now be proportional to the matrix elements of dimension-11 operators
\begin{eqnarray}\label{eq:mixed_factorization}
 x_D^{(11)} \sim  \sum_{q=s,d} C_{\bar{q}Gq} \langle D^0 | \left(\overline{c} \Gamma_1 u\right) \left(\overline{q} \sigma G q\right)\left(\overline{c}\Gamma_2 u\right) |\overline D^0\rangle  \sim \langle \overline{q} \sigma G q \rangle \sum_{q=s,d} C_{\bar{q}Gq} \langle D^0 | \left(\overline{c} \Gamma_1 u\right) \left(\overline{c}\Gamma_2 u\right) |\overline D^0\rangle \,.
 \nonumber \\
\end{eqnarray}

We obtain the mixed condensate amplitude analogously to the quark condensate as
\begin{equation}\label{eq:GqqStart}
T_{Gqq'}=\frac{iG_F^2}{2} \int d^4x \big(\overline{c}^i(x) \Gamma_\mu {S_1}(x, 0)^{j p} \Gamma^\nu u^q(0)\big)\big(\overline{c}^m(0) \Gamma_\nu\big)_\beta\big(q^n(0)\big)_\beta\big(\overline{q}^k(x)\big)_\alpha\big(\Gamma^\mu u^l(x)\big)_\alpha,
\end{equation}
where instead of the free propagator $S_0(x,y)$ we now pick up the contribution of the soft gluon field correction to the propagator $S_1(x,y)$, as seen in figure \ref{fig:mixed_condensate}. The soft gluon field is then factorized, together with the quark fields, into the mixed quark-gluon condensate. The propagator expansion for the light quark in the background gauge of Eq.~(\ref{PropExp}) has a term that involves the gluon field \cite{Novikov:1984ecy,Piscopo:2021ogu}, which in the coordinate representation can be written as
\begin{eqnarray}\label{eq:propGluon}
    S_1(x,0)^{jp} = &-&\frac{i}{8\pi^2}(\widetilde{G}_{\alpha\beta})^{jp}x^\alpha \gamma^\beta \gamma^5 \frac{m_{q'}}{\sqrt{-x^2}}K_1(m_{q'}\sqrt{-x^2})
    \nonumber \\
    &+& \frac{1}{16\pi^2}(G_{\alpha\beta})^{jp}\sigma^{\alpha\beta}m_{q'}K_0(m_{q'}\sqrt{-x^2}).
\end{eqnarray}

Expanding the nonlocal condensate of quark and gluon fields in \eqref{eq:mixed_factorization},  in terms of the local condensates, up to dimension-6 contribution, \cite{pascual2014qcd}, we obtain 
\begin{eqnarray}\label{eq:GqqExpansion}
&& \langle\overline{q}^{a}_{\alpha}(x) G_{\mu\nu}^{cd}(0)q_{\beta}^{b}(0)\rangle=\frac{1}{384}\left(\delta^{bd}\delta^{ac}-\frac{1}{3}\delta^{cd}\delta^{ab}\right) 
\\
&\times& \Bigg[ \big( \sigma_{\mu\nu}+m\left(i\sigma_{\mu\nu}\slashed{x}+\gamma_\mu x_\nu-\gamma_\nu x_\mu \right)\big)_{\beta\alpha}\lambda_q^2\langle\overline{q}q\rangle_0+\frac{i}{2}\left(\slashed{x}\sigma_{\mu\nu}\right)_{\beta\alpha}\frac{16}{9}\pi\alpha_s^{\rm NP}\langle\overline{q}q\rangle_0^2+\ldots\Bigg].
\nonumber
\end{eqnarray}
Note that it is precisely the terms proportional to $x_\mu$ in \eqref{eq:qq_cond} and $\sigma_{\mu\nu}x_\rho$ in \eqref{eq:GqqExpansion} that provide the necessary helicity flips to contribute to the mass difference mixing parameter while at the same time reducing powers of the $m_s/m_c$ GIM suppression. Also be aware that there is a higher dimensional, dimension-6 contribution present in the expansion, coming from the factorized four-quark condensate piece, $\langle\overline{q}q\rangle_0^2$ . Furthermore, it is worth pointing out the non-trivial color structure of the mixed condensate. 
Evaluating the integral in Eq.\eqref{eq:GqqStart} and taking matrix elements, we find the final result for the dimension-eleven operator
\begin{eqnarray}\label{eq:AnsDim11}
    x_D^{(11)} = && \frac{G_F^2}{2}\frac{\xi_s^2m_c^2}{2M_D\Gamma_D}\bigg(\frac{\lambda_q^2}{m_c^2}\frac{\langle\overline{s}s\rangle_0}{m_c^3}x_s^3 + \frac{8}{9}\pi\alpha_s^{NP}\frac{\langle\overline{s}s\rangle_0^2-\langle\overline{d}d\rangle_0^2}{m_c^6} x_s^2 \bigg)
    \nonumber \\
    &\times&
    \frac{1}{6}\big( C_1^2 \langle O_{V-A} \rangle+8 \langle O_{S-P} \rangle(C_1^2+4C_1C_2+6C_2^2) \big).
\end{eqnarray}
The first term  has the same parametric $m_s^3$ dependence on $m_s$, as the quark-quark condensate contribution in $x_D^{(9)}$, see Eq.\eqref{AnsDim9}. However, here the parametrically leading $m_s^2$ dependence comes from the higher-dimensional four-quark condensates, $\sim \langle \overline{q} q \rangle_0^2$ terms, but these terms are additionally suppressed by $1/m_c^3$ and by the $SU(3)_F$-breaking stemming from the difference of $\langle \bar s s\rangle_0$ and $\langle \bar d d\rangle_0$ condensates. We will discuss the numerical value of the $x_D^{(11)}$ contribution in Section \ref{sec:Summary}.
%

%%%%%%%%%%%%%%%%%%%%%%%%%%%%%%%%%%%%%%%%%%
\subsection{Dimension-twelve operator contributions}
\label{sec:qqqq}

As it was emphasized in Section \ref{sec:qq}, factorization of dimension-nine matrix elements allowed us to replace one of the mass insertions with the quark condensate, which also plays a role of the QCD order parameter and flips the chirality of the light $s$-quark, which reduces the GIM suppression to $\mathcal{O}(m_s^3/m_c^3)$. In order to get this contribution, we "cut" the fermion propagator -- or, look at the infrared part of the propagator contribution.

\begin{figure}
    \centering
\includegraphics[width=0.5\linewidth]{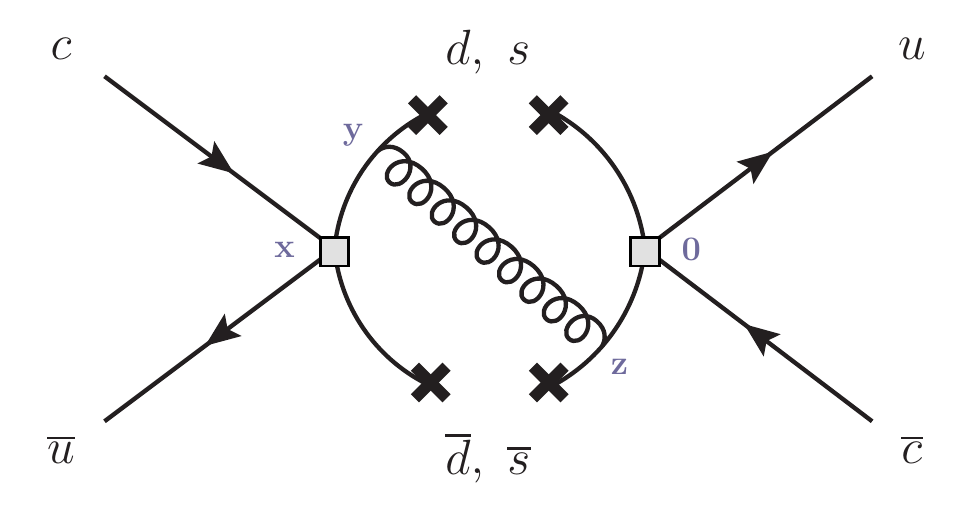}
       \caption{An example of a diagram of the dimension-12 contribution proportional to the four quark  condensates \eqref{eq:4qfactorization}, where the gluon connects opposite internal quark lines. Note that there are sixteen diagrams in total, since each of the four lines on the left can connect to any of the four lines on the right. This makes the color structure of the separate contributions nontrivial.}
    \label{fig:4q_int-intPrime_xyz}
\end{figure}

In principle, the same can be done with the other propagator. Naively, this should remove another mass insertion, making the overall contribution to scale as $\mathcal{O}(m_s^2/m_c^2)$, which will make it consistent with the order of the $SU(3)_F$ symmetry breaking obtained by general group-theory methods \cite{Falk:2001hx}. However, one needs to transfer momentum through the diagram, which implies that a perturbative QCD correction is required: the gluon propagator can serve such purpose, as shown in Fig. \ref{fig:4q_int-intPrime_xyz}. The resulting contribution is that of dimension-12 operators, further suppressed by powers of $1/m_c$, but having less powers of $m_s$. Extending the same logical approach as in previous sections, we can again factorize the resulting operator into a four-quark vacuum condensate and a matrix element of dimension-6 operators. In this section for the first time we obtain explicitly the contribution of the four-quark condensate and confirm it is parametrically leading, i.e. has mass dependence $\mathcal{O}(m_s/m_c)^2$ as predicted in \cite{Falk:2001hx,Georgi:1992as}. 

We obtain the four-quark contribution by factorizing all four internal quarks using vacuum insertion approximation. The four legs on either side can be connected in sixteen different ways, an example of which is shown in Fig. \ref{fig:4q_int-intPrime_xyz} and its amplitude reads
\begin{eqnarray}
T_{4q}=i \int d^4x\, d^4y\, d^4z\, (-ig)^2\big({\overline{q}_1^m}(y) \gamma^\rho (T^a)^{mn} S_{q_1}^{ni}(y,x) \Gamma^\mu c^j(x)\big)\big(\overline{u}^k(x) \Gamma_\mu {q_3^l}(x)\big) \\
\times S_G(y,z)\big(\overline{u}^q(0) \Gamma_\nu {q_2^r}(0)\big)\big({\overline{q}_4^o}(z) \gamma_\rho (T^a)^{op}  S_{q_4}^{ps}(z,0) \Gamma^\nu c^t(0)\big),
\nonumber
\end{eqnarray}
where $S_G(y,z)^{ab}_{\rho\sigma}=\delta^{ab}g_{\rho\sigma}S_G(y,z)$ is the gluon propagator, the $q_{1,2,3,4}$ indicate the four internal quarks, and $S^{ij}_{q}(x,y)=\delta^{ij}S_{q}(x,y)$ are their respective propagators. Factorizing, we obtain
\begin{equation}
\begin{aligned}
T_{4q}=(T^a)^{mn}(T^a)^{op}\delta^{ni}\delta^{ps}i \int d^4x\, d^4y\, d^4z\, 
(-ig)^2\langle {{\overline{q}_1^m}(y)}_\alpha {{\overline{q}_4^o}(z)}_\delta {{q_3^l}(x)}_\gamma {{q_2^r}(0)}_\beta\rangle& \\
\times\big( \gamma^\rho  S_{q_1}(y,x) \Gamma^\mu c^j(x)\big)_\alpha\big(\overline{u}^k(x) \Gamma_\mu \big)_\gamma S_G(y,z)\big(\overline{u}^q(0) \Gamma_\nu \big)_\beta \big( \gamma^\rho S_{q_4}(z,0)& \Gamma^\nu c^t(0)\big)_\delta,
\end{aligned}
\end{equation}
where $\alpha,\, \beta,\, \gamma,\, \delta$ are Dirac indices. The value of the four-quark condensate has not yet been determined model-independently. In order to get the numerical estimate of this contribution, we follow the approach in Ref. \cite{pascual2014qcd} and further factorize this contribution into a product of two two-quark vacuum condensates\footnote{A previous attempt at computing the four-quark contribution claimed there were problems with second factorization \cite{Yeghiyan:2012vq}. In our calculation we do not encounter any issues in the factorization process.}, 
\begin{equation}\label{eq:4qfactorization}
\langle {{\overline{q}_1^m}(y)}_\alpha {{\overline{q}_4^o}(z)}_\delta {{q_3^l}(x)}_\gamma {{q_2^r}(w)}_\beta\rangle = \langle {{\overline{q}_1^m}(y)}_\alpha{{q_2^r}(w)}_\beta\rangle\langle{{\overline{q}_4^o}(z)}_\delta{{q_3^l}(x)}_\gamma\rangle - \langle {{\overline{q}_1^m}(y)}_\alpha{{q_3^l}(x)}_\gamma\rangle \langle{{\overline{q}_4^o}(z)}_\delta{{q_2^r}(w)}_\beta\rangle,
\end{equation}
where the two terms represent the four-quark condensate forming from different pairs of quarks. The first term corresponds to the upper and lower pairs of quarks forming the condensates, while the second corresponds to the left and right pairs forming condensates. The second term, naturally, only contributes if both lines are of the same flavor.

The richness of structure of these amplitudes is apparent in color space as well. The Gell-Mann matrices at each QCD vertex, coupled with distinct color flows through condensates and the different contractions from the weak $Q_1$ and $Q_2$ operators, collectively yield a result whose numerical value is challenging to predict without explicit computations due to substantial variations in magnitude (and even sign) of each contribution. Some simplifications are possible by using symmetry relations between some of the diagrams (resulting from CP symmetry relations). For example the diagram in which the gluon connects the incoming $c$ quark with the upper internal quark is equivalent to the diagram in which the gluon connects the lower internal quark line to the outgoing $c$ quark.  

We present the results diagram-by-diagram, in Appendix \ref{app:4qdiag-by-diag}, in order to better understand the interplay of enhancement from strong $SU(3)_F$ breaking, and suppression from this contribution being higher order in the OPE. In the results presented, only terms leading in powers of $m_s/m_c$ are shown; the Wilson coefficients and operators are evaluated at a renormalization scale $\mu=1.3$ GeV, with other parameters listed in Section \ref{sec:Summary}. The final, total, result for the four-quark condensate reads
\begin{equation}
\begin{aligned}
x_D^{(12)}=\frac{G_F^2}{2}\frac{\xi_s^2m_c^2}{2M_D\Gamma_D}\frac{\langle\overline{s}s\rangle_0^2}{m_c^6}\pi\alpha_s(\mu) x_s^2&\\
\times\frac{1}{108}\bigg[ 3(-9 C_1^2+172 &C_1 C_2+411C_2^2)\langle O_{V-A} \rangle \\
&+ 8(13 C_1^2+452 C_1 C_2+495 C_2^2)\langle O_{S-P} \rangle
\bigg].
\end{aligned}
\end{equation}
The expected mass dependence $m_s^2$ in the final results arises from various terms involving the interplay of helicity flips caused by propagator mass insertions and the condensate. For example, in the diagram shown in Fig. \ref{fig:4q_int-intPrime_xyz}, each line can feature $SU(3)_F$ breaking terms resulting from two helicity flips—one from a single mass insertion in the propagator, and the other from the condensate. Alternatively, both helicity flips could come from the mass-dependent part of the condensate, eliminating the need for any mass insertions from the propagator. Each sequence of helicity flips forms a separate term that contributes to the final outcome. Note that the gluon connects the diagram in sixteen different ways, allowing for other combinations of helicity flips in different diagrams. A noteworthy observation is that the presented dimension-12 contribution is proportional to the perturbative coupling $\alpha_s(\mu)$, whereas the dimension-12 terms in the local expansion of the nonlocal quark–quark condensate \eqref{eq:qq_cond} and the nonlocal mixed quark–gluon condensate \eqref{eq:GqqExpansion} are proportional to the nonperturbative coupling $\alpha_s^{\mathrm{NP}}$.

Numerically, contrary to earlier estimates in the literature, we do not find the contribution from dimension-12 operators numerically leading. In fact, the combined result falls well below the experimental value. We attribute this discrepancy to taking the condensate value as the `hadronic scale' $\Lambda=1$ GeV in back-of-the-envelope estimates of $\langle\overline{q}q\rangle_0= -\Lambda^3$. In reality, the relevant scale is that of the quark condensate, which has a known value of $\Lambda\approx 0.3$ GeV. The lower actual value compared to previous estimates has a substantial impact, especially on the four-quark condensate contribution, since it is proportional to $\langle\overline{q}q\rangle_0^2=\Lambda^6$. In simple terms, although the charm quark's mass is not heavy enough to strongly suppress higher-order terms in the OPE, the value of the condensate is small enough that \textit{it} provides the suppression itself. The contribution is further suppressed due to mutual cancellations between different terms (see results in Appendix \ref{app:4qdiag-by-diag}).

%%%%%%%%%%%%%%%%%%%%%%%%%%%%%%%%%%%%%%%%%%%%%%%%%%%%%
\section{Summary}
\label{sec:Summary}

It would be helpful to analyze both the numerical value of the total answer and its uncertainty. We summarize our results in the table \ref{tab:finalResults}. 
\begin{table}
    \centering
    \renewcommand{\arraystretch}{1.4} % Increase row spacing
    \begin{tabular}{l c}
        \textbf{Contribution} & \textbf{Result for $x_D$} [$10^{-5}$] \\
        $x_D^{(9)}$        & $0.99 \pm 0.13$ \\
        $x_D^{(11)}$                  & $0.051 \pm 0.009$ \\
        $x_D^{(12)}$         & $0.24 \pm 0.06$ \\\hline\hline
        Total $= x_D^{\rm cond}$  & $1.27 \pm 0.18$ \\\hline\hline
        $x_D^{LO}$      & $0.37 \pm 0.02$ \\
        $x_D^{NLO}$ \cite{Golowich:2005pt} & $-0.45 \pm 0.02$ \\
    \end{tabular}
    \caption{The contributions to $x_D$ from different orders in $1/m_c$, and $\alpha_s$ expansions.}
    \label{tab:finalResults}
\end{table}
While the contributions from dimension-12 operators (four-quark condensate) are parametrically leading with only two powers of $x_s$, they are not necessarily the largest in magnitude. We find that the contributions from the dimension-6, dimension-9, and dimension-12 operators are approximately the same in size. Additionally, we observe that the contribution $x_D^{(11)}$, being proportional to the  mixed quark-gluon condensate, also contains  higher-dimensional four-quark terms from the local expansion of the condensate which are proportional to $x_s^2$. However, numerically, the $x_D^{(11)}$ is about an order of magnitude smaller than $x_D^{(9)}$ and even $x_D^{(12)}$, due to accidental numerical suppression in the second line of \eqref{eq:AnsDim11}.

\begin{figure}
    \centering
    \includegraphics[width=0.5\linewidth]{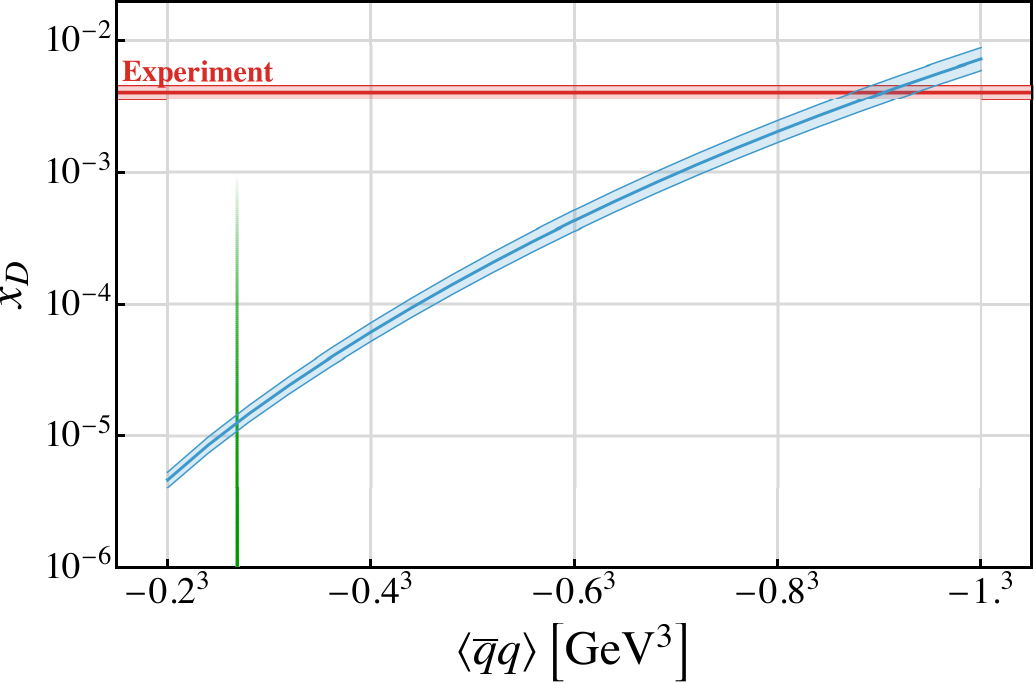}
    \caption{Dependence on the value of the quark condensate illustrates how our result can reach the experimental order of magnitude if taken to be around the hadronic scale $\approx 1$ GeV$^3$ as was previously estimated \cite{Bigi:2000wn}. Unfortunately, the actual value of the condensate (indicated by the green line) is significantly smaller than $\approx 1$ GeV$^3$, see table \ref{tab:numericalvalues}.}
    \label{fig:conddep}
\end{figure}

The NLO contribution is taken from \cite{Golowich:2005pt}, using the input parameters and scale employed in this paper\footnote{Note that the choice of the overall sign must be such that the final result is positive, since the $x_D$ parameter represents the mass difference. That is why, in \cite{Golowich:2005pt}, the NLO contribution is positive and the LO is negative, whereas in this work it is the opposite.}. We present our results with errors estimated from uncertainties in input parameters. Results are calculated at the renormalization scale $\mu=1.3$ GeV, with other input parameters listed in the appendix. 
\begin{figure}
    \centering
    \includegraphics[width=0.5\linewidth]{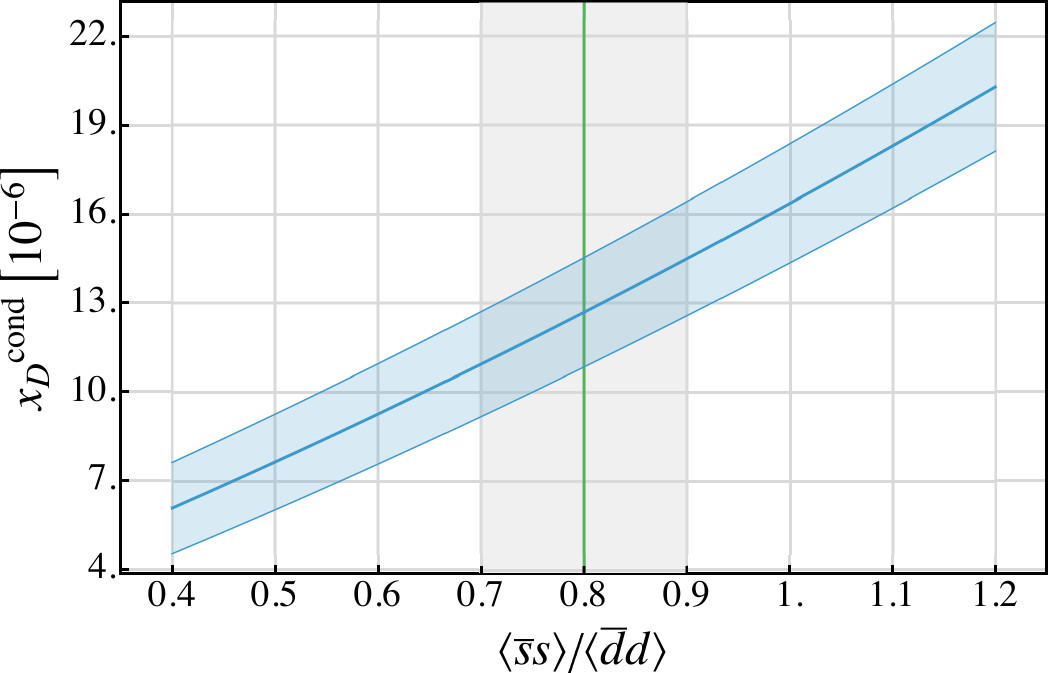}
    \caption{We observe a strong dependence of our result on the relatively unknown $SU(3)_F$ breaking parameter $\langle\overline{s}s\rangle_0/\langle\overline{d}d\rangle_0$, with the cental value $\langle\overline{s}s\rangle_0/\langle\overline{d}d\rangle_0 = 0.8 \pm 0.1$.}
    \label{fig:Rsddep}
\end{figure}
\begin{figure}
    \centering
    \includegraphics[width=0.57\linewidth]{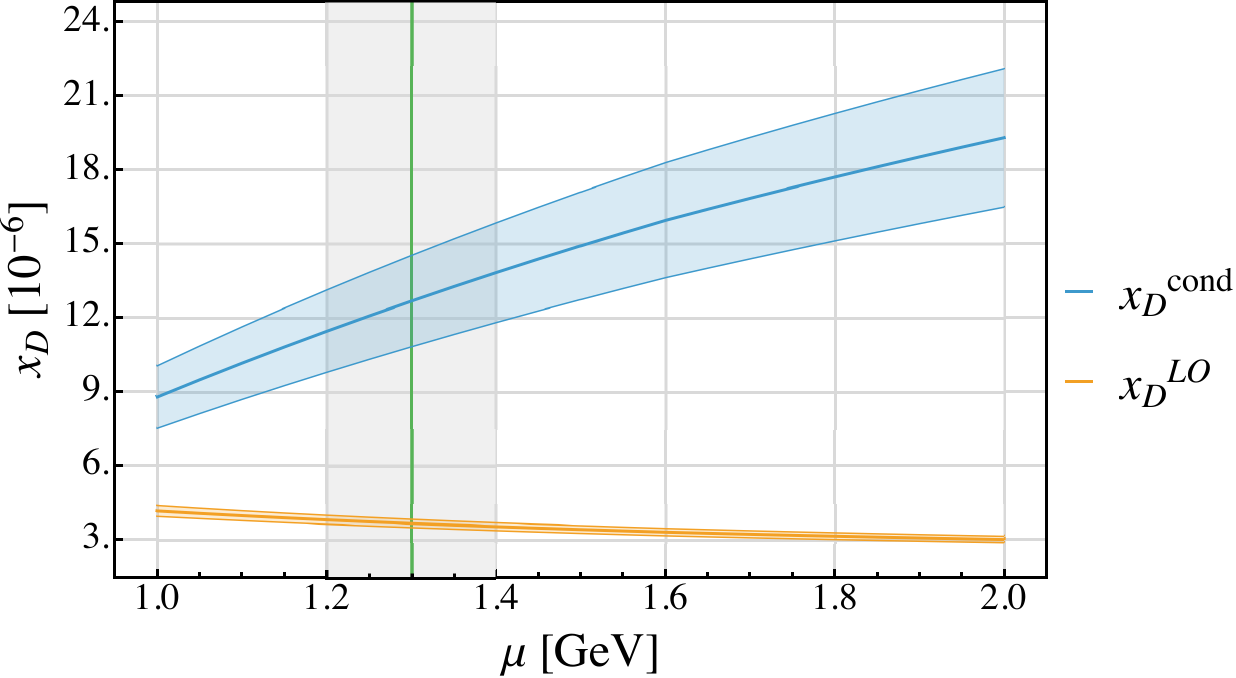}
    \caption{A strong dependence on renormalization scale, taken at the central value $\mu = 1.3 \pm 0.1$ GeV, is expected since the condensate is strongly scale dependent. The leading order perturbative result is shown for comparison.}
    \label{fig:mudep}
\end{figure}
We evaluated possible uncertainties by varying parameters of the computation and then adding them in quadrature, as it often done for theoretical calculations. We present the most important sources of uncertainties in Fig.~\ref{fig:Rsddep} (the $SU(3)_F$ breaking in the ratio of $\langle\overline{s}s\rangle_0/\langle\overline{d}d\rangle_0$ condensates for which we take the central value $\langle\overline{s}s\rangle_0/\langle\overline{d}d\rangle_0=0.8 \pm 0.1$ \cite{Reinders:1984sr}), Fig.~\ref{fig:mudep} (the dependence on renormalization scale $\mu$), and Fig.~\ref{fig:conddep} (the numerical value of the condensate).
There are other sources of uncertainty not discussed in the paper, which relate to the terms in the OPE that we have not considered here, specifically those that do not cause a chirality flip for the light quarks. Such terms have the same powers of $m_s$, but are additionally suppressed by more powers of $1/m_c$. Therefore, they are unlikely to produce numerically larger effects in the prediction of $x_D$.

%%%%%%%%%%%%%%%%%%%%%%%%%%%%%%%%%%%%%%%%%%%%%%%%%%%%%
\section{Conclusions}
\label{sec:Conclusions}

We calculated the $\ddbar$ mixing parameter $x_D$ using the operator product expansion formalism in QCD, expanding the relevant correlation functions in powers of $1/m_c$. Following earlier studies \cite{Georgi:1992as,Golowich:2005pt,Bigi:2000wn,Bobrowski:2009zc}, we focused only on the terms that produce the fewest powers of $m_s$, even though such terms might be suppressed by higher powers of $1/m_c$, calculating consistently terms up to dimension-12 operators. In addition to the leading-order contribution from dimension-6 operators, we explicitly calculated matrix elements of operators with dimensions 9, 11, and 12, whose contributions, in factorization approximation, are proportional to the two-quark, mixed quark-gluon, and four-quark QCD condensates. Contrary to the previous estimates, our explicit calculations does not find a large enhancement from such higher-order terms. The numerical result,
\beq
x_D^{\rm cond}=  (1.27 \pm 0.18) \times 10^{-5}
\eeq
is about two  orders of magnitude lower than the experimental value \cite{HeavyFlavorAveragingGroupHFLAV:2024ctg}, 
\beq
x_D^{\rm exp} = (0.407 \pm 0.044) \%. 
\eeq
This could be traced to the fact that previous studies overestimated the value of the quark-quark condensate.

It might be possible that partial resummations of various condensate contributions provide some boost to the predicted value \cite{Dulibic:2024lse}. Yet, a complete computation of relevant matrix elements, without the assumption of factorization, is desired. Alternatively, a direct computation of the mixing matrix elements on the lattice could soon provide a new estimate of the $\ddbar$ mixing parameters \cite{DiCarlo:2025mvt}.

%%%%%%%%%%%%%%%%%%%%%%%%%%%%%%%%%%%%%%%%%%%%%%%%%
\section*{Acknowledgments}
The authors acknowledge the support of the Croatian Science Foundation (HRZZ) under project “Nonperturbative QCD in heavy flavour physics” (IP-2024-05-4427). AAP was supported in part by the US Department of Energy grant DE-SC0024357. This work was performed in part at Aspen Center for Physics, which is supported by National Science Foundation grant PHY-2210452.

%%%%%%%%%%%%%%%%%%%%%%%%%%%%%%%%%%%%%%%%%%%%%%%%%
\appendix

%%%%%%%%%%%%%%%%%%%%%%%%%%%%%%%%%%%%%%%%%%%%%%%%%

\section{Leading order expressions}\label{app:integral}

Here we explicitly give functions $X(m_1^2,m_2^2,p^2)$ and $Y(m_1^2,m_2^2,p^2)$ used in presenting the leading order result \eqref{eq:LO_intermediate-stepXY}:
\begin{equation}
\begin{split}
    X(m_1^2,m_2^2,p^2)=\frac{-i}{{192 \pi^2 p^2}} \bigg( \left( m_1^4 - 2 m_1^2 \left( m_2^2 + p^2 \right) + \left( m_2^2 - p^2 \right)^2 \right) B_0(p^2, m_1^2, m_2^2) +\\
    +A_0(m_1^2)(- (m_1^2 - m_2^2 + p^2)) + A_0(m_2^2)(m_1^2 - m_2^2 - p^2) \bigg)\,,
\end{split}
\end{equation}
\begin{equation}
\begin{split}
    Y(m_1^2,m_2^2,p^2)=\frac{i}{96 \pi^2 p^4} \bigg( \left( 2 m_1^4 - m_1^2 \left( 4 m_2^2 + p^2 \right) + 2 m_2^4 - m_2^2 p^2 - p^4 \right) B_0(p^2, m_1^2, m_2^2) +\\
    +A_0(m_1^2) \left( -2 m_1^2 + 2 m_2^2 + p^2 \right) + A_0(m_2^2) \left( 2 m_1^2 - 2 m_2^2 + p^2 \right) \bigg) \,.
\end{split}
\end{equation}
The functions $A_0$ and $B_0$ are standard one-point and two-point scalar functions in perturbative calculations, respectively. 

%%%%%%%%%%%%%%%%%%%%%%%%%%%%%%%%%%%%%%%%%%%%%%%%%
\section{Four-quark condensate contributions from Sec. \ref{sec:qqqq}}
\label{app:4qdiag-by-diag}
In this appendix we provide analytical (and numerical) results for all the independent diagrams which together form the four-quark condensate contribution. There are sixteen diagrams in total, of which nine are independent, while the rest can be deduced through symmetry arguments. 

Note how the different combinations of Wilson coefficients give rise to factors varying in magnitude, and even sign, making the contributions sometimes cancel among themselves in the final result.

\begin{equation}\label{eq:4q_intintPrime}
\begin{aligned}
    x_D^{(12)}\left(\raisebox{-15pt}{\includegraphics[width=65pt]{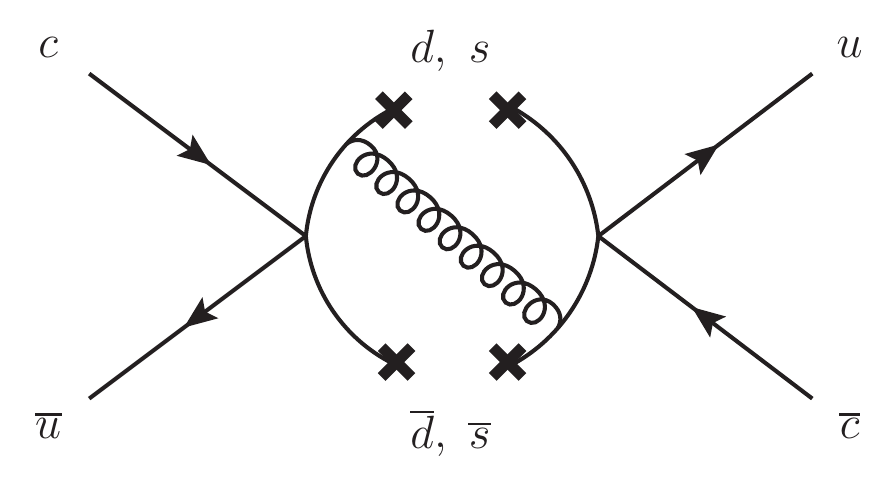}}\right)& \approx -7\times 10^{-8} = - \frac{4 G_F^2}{2} \xi^2 \frac{m_c^2}{2\Gamma_D M_D}\frac{4\pi\alpha_s}{108}\frac{\langle\overline{s}s\rangle_0^2}{m_c^6}x_s^2 \\
    &\times {\Big[ \left(C_1^2 + 4C_1C_2 + 6C_2^2\right)  \langle O_{V-A} \rangle + 4 \left(2 C_1C_2 + 3 C_2^2 \right)   \langle O_{S-P} \rangle \Big]}
    \nonumber
\end{aligned}
\end{equation}
\begin{equation}\label{eq:4q_intint}
\begin{aligned}
    x_D^{(12)}\left(\raisebox{-15pt}{\includegraphics[width=65pt]{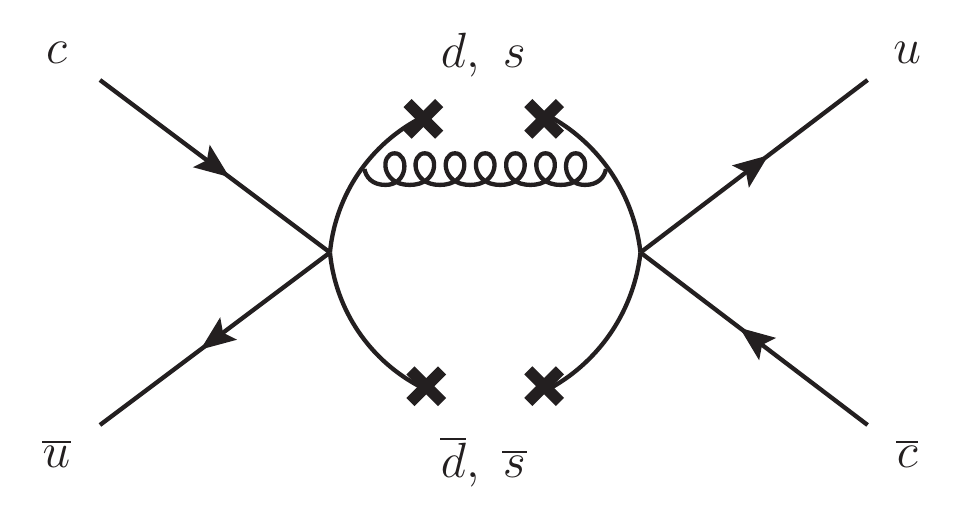}}\right)&\approx +2\times 10^{-6} = - \frac{G_F^2}{2} \xi^2 \frac{m_c^2}{2\Gamma_D M_D}\frac{4\pi\alpha_s}{108}\frac{\langle\overline{s}s\rangle_0^2}{m_c^6}x_s^2 \phantom{qwertytyu} \\
    & \times \Big[ \left( 27 C_1^2-112 C_1 C_2-168 C_2^2\right)  \langle O_{V-A} \rangle  \\
    & \qquad + 8 \left(22 C_1^2 -42 C_1C_2 + 63 C_2^2 \right)   \langle O_{S-P} \rangle  \Big],
    \nonumber
\end{aligned}
\end{equation}
\begin{equation}\label{eq:4q_extCint}
\begin{aligned}
    x_D^{(12)}\left(\raisebox{-15pt}{\includegraphics[width=65pt]{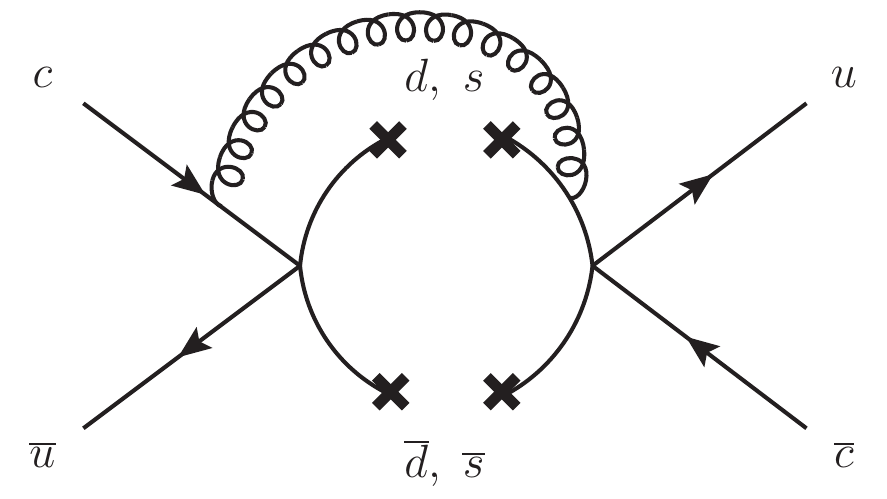}}\right)&\approx -7\times 10^{-7} = \frac{7 G_F^2}{4}\xi^2 \frac{m_c^2}{2\Gamma_D M_D}\frac{4\pi\alpha_s}{108}\frac{\langle\overline{s}s\rangle_0^2}{m_c^6}x_s^2 \phantom{qwertytyuuu}  \\
    &\times \Big[ 
    -7 C_1C_2 \langle O_{V-A} \rangle +
    8 \left( 2 C_1^2 - C_1C_2\right)  \langle O_{S-P} \rangle
    \Big],
    \nonumber
\end{aligned}
\end{equation}
\begin{equation}\label{eq:4q_extCintPrime}
\begin{aligned}
    x_D^{(12)}\left(\raisebox{-15pt}{\includegraphics[width=65pt]{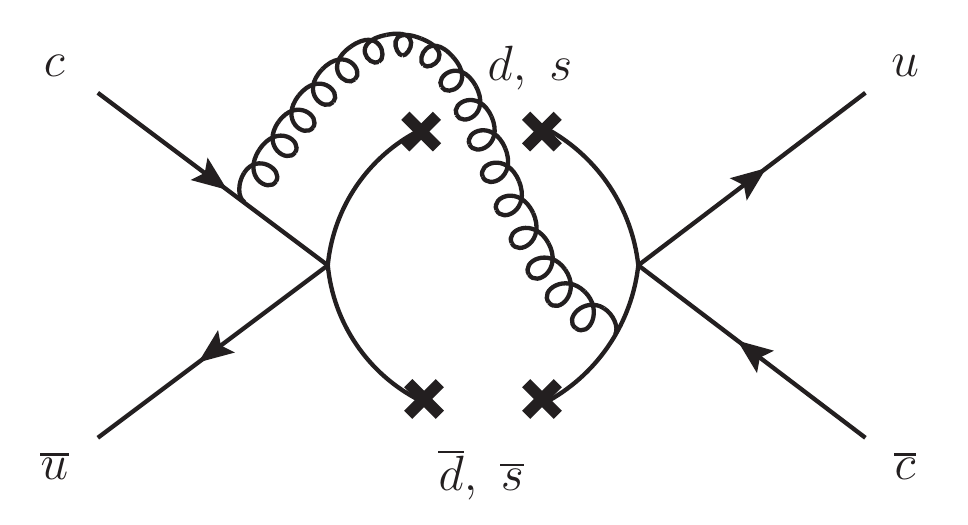}}\right)&\approx +2\times 10^{-7} =\frac{G_F^2}{2}\xi^2 \frac{m_c^2}{2\Gamma_D M_D}\frac{4\pi\alpha_s}{108}\frac{\langle\overline{s}s\rangle_0^2}{m_c^6}x_s^2 \phantom{qwertytyuuu} \\
    &\times \Big[ \left(3 C_1^2 + C_1C_2\right) \langle O_{V-A} \rangle
    + 28 \left(C_1^2 + 3 C_1C_2\right) \langle O_{S-P} \rangle \Big], \nonumber
\end{aligned}
\end{equation}
\begin{equation}\label{eq:4q_extUintPrime}
\begin{aligned}
    x_D^{(12)}\left(\raisebox{-15pt}{\includegraphics[width=65pt]{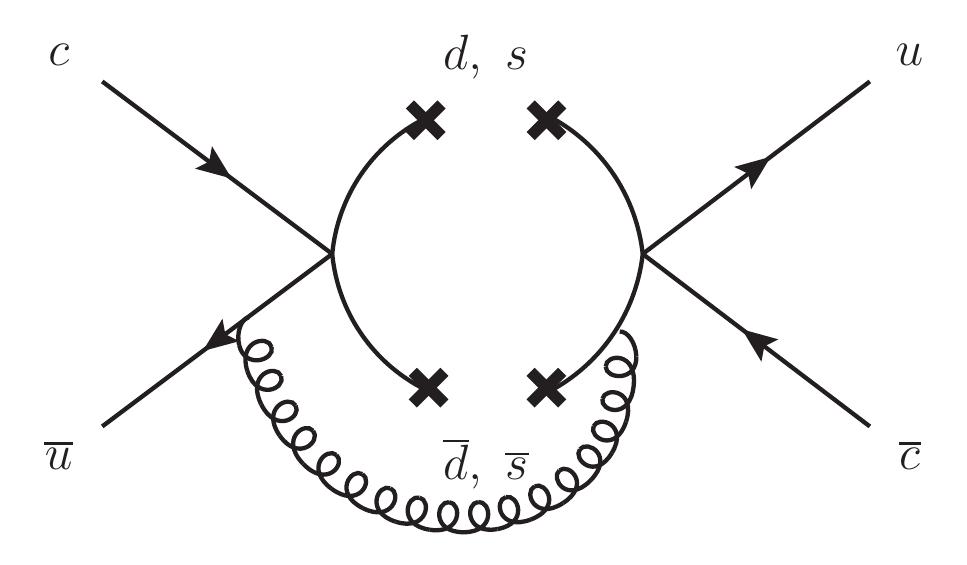}}\right)&\approx -6\times 10^{-7} =\frac{G_F^2}{2}\xi^2 \frac{m_c^2}{2\Gamma_D M_D}\frac{4\pi\alpha_s}{108}\frac{\langle\overline{s}s\rangle_0^2}{m_c^6}x_s^2 \nonumber \\
    &\times\Big[ 
    C_1 \left(16 C_1-C_2\right) \langle O_{V-A} \rangle
    + 24 C_1 \left(2 C_1-C_2\right) \langle O_{S-P} \rangle
    \Big],
\end{aligned}
\end{equation}
\begin{equation}\label{eq:4q_extUint}
\begin{aligned}
    x_D^{(12)}\left(\raisebox{-15pt}{\includegraphics[width=65pt]{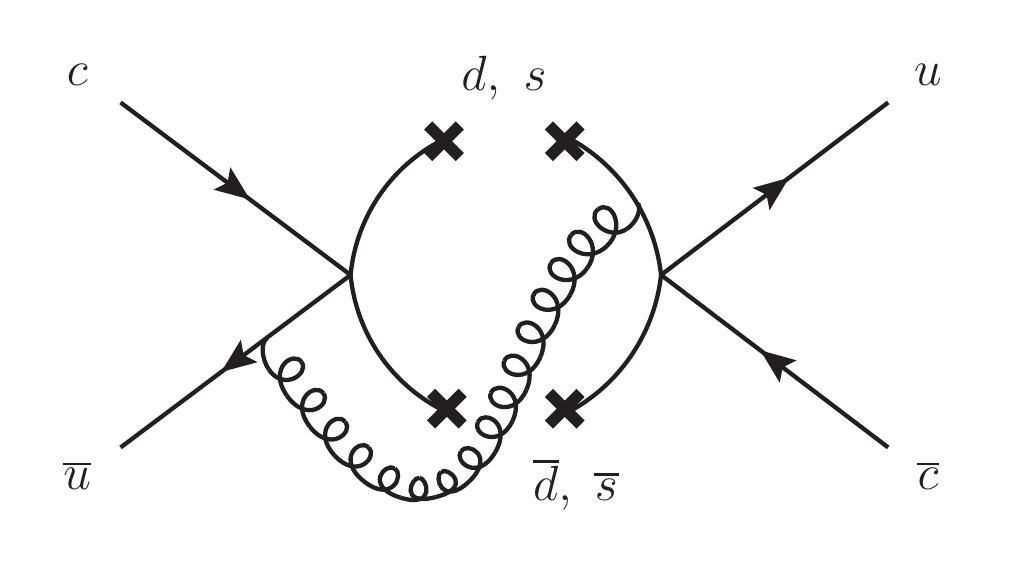}}\right)&\approx -3\times 10^{-9} =\frac{G_F^2}{8} \xi^2 \frac{m_c^2}{2\Gamma_D M_D}\frac{4\pi\alpha_s}{108}\frac{\langle\overline{s}s\rangle_0^2}{m_c^6}x_s^2 \nonumber \\
    &\times \Big[ 
    - C_1 \left(7 C_1+59 C_2\right) \langle O_{V-A} \rangle
    + 24 C_1 \left(C_1+C_2\right) \langle O_{S-P} \rangle
    \Big],
\end{aligned}
\end{equation}
\begin{equation}\label{eq:4q_extCextU}
    \begin{aligned}
        x_D^{(12)}&\left(\raisebox{-15pt}{\includegraphics[width=65pt]{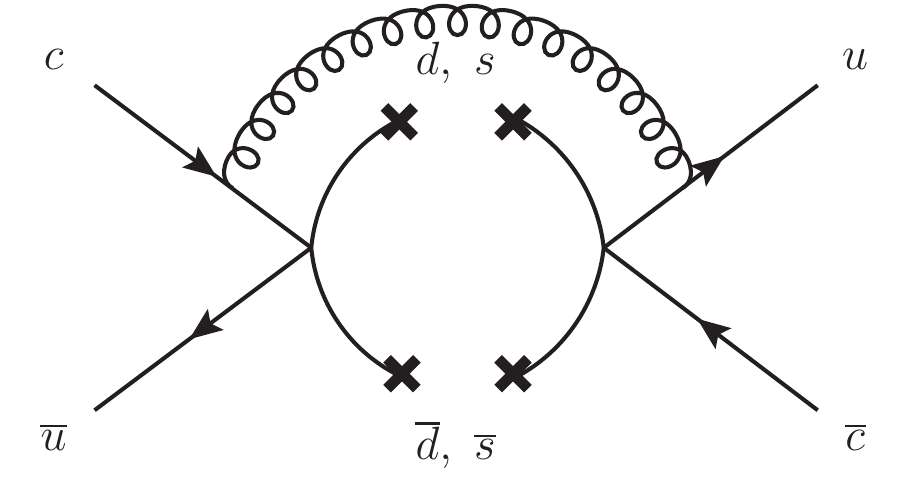}}\right)\approx -2\times 10^{-7} =\frac{G_F^2}{4} \xi^2 \frac{m_c^2}{2\Gamma_D M_D}\frac{4\pi\alpha_s}{108}\frac{\langle\overline{s}s\rangle_0^2}{m_c^6}x_s^2 \nonumber \\
        &\times\Big[ 
        \left(16 C_1^2+14 C_1 C_2+21 C_2^2 \right) \langle O_{V-A} \rangle
        + 24 \left(2 C_1^2+2 C_1 C_2+3 C_2^2\right) \langle O_{S-P} \rangle
        \Big]
    \end{aligned}
\end{equation}
\begin{equation}\label{eq:4q_extCextC}
    \begin{aligned}
        x_D^{(12)}&\left(\raisebox{-15pt}{\includegraphics[width=65pt]{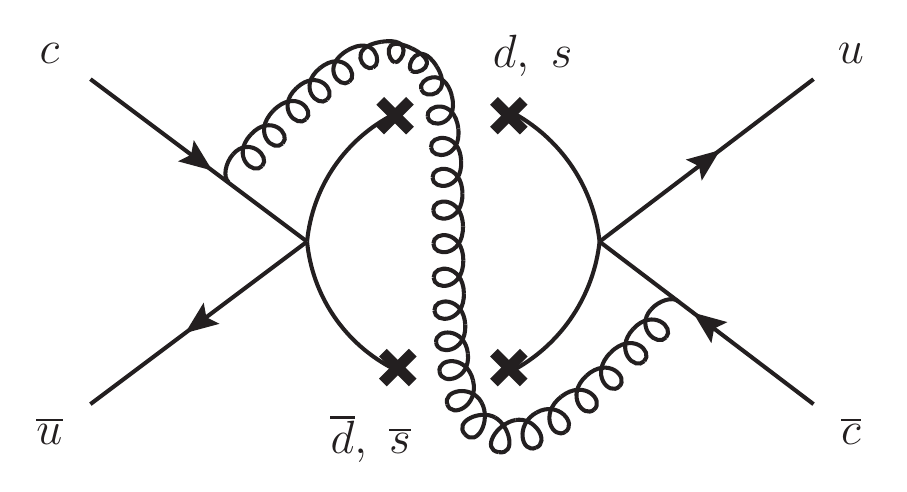}}\right)\approx +2\times 10^{-7} = - \frac{G_F^2}{8} \xi^2 \frac{m_c^2}{2\Gamma_D M_D}\frac{4\pi\alpha_s}{108}\frac{\langle\overline{s}s\rangle_0^2}{m_c^6}x_s^2 \nonumber \\
        &\times \Big[
        \left(19 C_1^2-18 C_1 C_2-27 C_2^2\right) \langle O_{V-A} \rangle
        + 56 (C_1-3 C_2) (C_1+C_2) \langle O_{S-P} \rangle
       \Big],
    \end{aligned}
\end{equation}
\begin{equation}\label{eq:4q_extUextU}
    \begin{aligned}
        x_D^{(12)}&\left(\raisebox{-15pt}{\includegraphics[width=65pt]{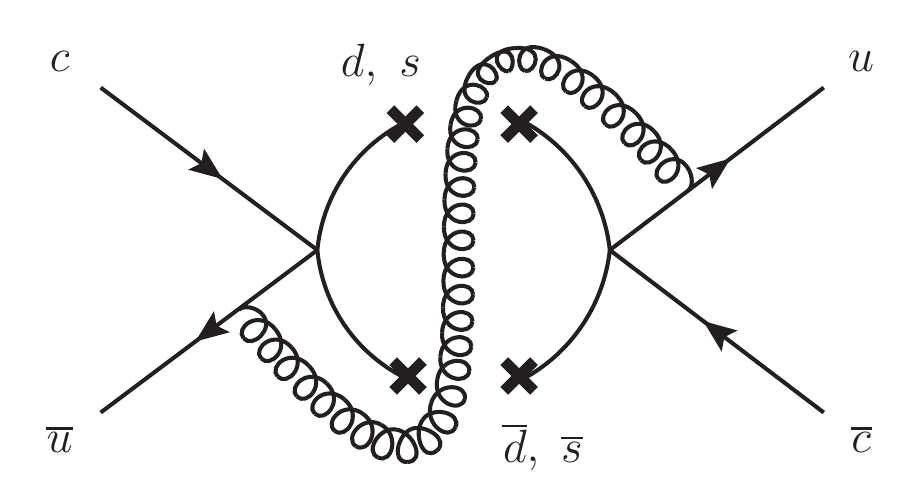}}\right)\approx -2\times 10^{-7} =\frac{G_F^2}{4}\xi^2\frac{m_c^2}{2\Gamma_D M_D}\frac{4\pi\alpha_s}{108}\frac{\langle\overline{s}s\rangle_0^2}{m_c^6}x_s^2 \nonumber \\
        &\times \Big[
        \left(7 C_1^2-10 C_1 C_2-15 C_2^2\right) \langle O_{V-A} \rangle
        + 24 (C_1-3 C_2) (C_1+C_2) \langle O_{S-P} \rangle
       \Big].
    \end{aligned}
\end{equation}

%%%%%%%%%%%%%%%%%%%%%%%%%%%%%%%%%%%%%%%%%%%%%%%%%
\section{Numerical input}

Standard parameters are taken from \cite{ParticleDataGroup:2024cfk}. The rest of the parameters are cited in Table \ref{tab:numericalvalues}. 

\renewcommand{\arraystretch}{1.5}
\begin{table}[th]
    \centering
    \begin{tabular}{c|c}
        Parameter & Numerical value \\ \hline\hline
        $f_D$ \cite{FlavourLatticeAveragingGroupFLAG:2024oxs}& $0.212\,\mathrm{GeV}$ \\
        $\langle O_{V-A}\rangle(3\,\mathrm{GeV})$ \cite{Bazavov:2017weg}& $(0.322\pm0.022)$ GeV$^4$ \\
        $\langle O_{S-P}\rangle(3\,\mathrm{GeV})$ \cite{Bazavov:2017weg}& $(-0.624\pm0.007)$ GeV$^4$\\
        $\langle\overline{q}q\rangle_0(1.3 \,\mathrm{GeV})$& $(-268.5\pm 1.3\,\mathrm{MeV})^3$ 
    \end{tabular}
    \caption{Numerical values of the decay constant and matrix elements used in our calculations. The condensate value is obtained via the Gell-Mann-Oakes-Renner relation.}
    \label{tab:numericalvalues}
    \end{table}

\bibliography{refs}% Produces the bibliography via BibTeX.
\bibliographystyle{JHEP}

\end{document}